\documentclass[electronics,article,submit,pdftex,oneauthor]{Definitions/mdpi}

\firstpage{1} 
\pubvolume{1}
\issuenum{1}
\articlenumber{0}
\pubyear{2026}
\copyrightyear{2026}
\datereceived{ } 
\daterevised{ } 
\dateaccepted{ } 
\datepublished{ } 

\usepackage[english]{babel}
\usepackage[nottoc]{tocbibind}
\usepackage{comment}
\usepackage{booktabs}
\usepackage{pgfplotstable}
\pgfplotsset{compat=1.18}
\usepackage{afterpage}
\usepackage{rotating}
\usepackage{multicol}
\usepackage{fancyvrb}
\usepackage{verbatim}
\newcommand{\fsl}{\textsf{FSL}}
\newcommand{\sle}{\textsf{SLE}}
\newcommand{\yas}{\textsf{YAS}}
\newcommand{\ontoeng}{\textsf{OE}}
\newcommand{\genai}{\textsf{GenAI}}

\Title{Towards an Ontology for the Foundations of Software Languages%
  \mbox{}
  \\
  \textbf{\color{red}\large{}Draft as of \today. Submitted for review/publication.}
}

\Author{Ralf L\"ammel $^{1}$\orcidA{}}

\AuthorNames{Ralf L\"ammel}

\address[1]{%
$^{1}$ \quad University of Koblenz, Faculty of Comuter Science, SoftLang Team; laemmel@uni-koblenz.de}

\abstract{%
  The notion of software languages subsumes programming languages,
  modeling languages, and yet many other types of languages used in
  software engineering. The emerging ontology `Foundations of Software
  Languages' (\fsl) organizes the foundations underlying software
  languages. We are concerned with language categories, language
  concepts, associated tools and methodological approaches, the formal
  systems or other formal entities underlying software languages, and
  the embedding of software languages into into software engineering
  activities. The primary objective of \fsl{} is to serve as a
  knowledge resource in Computer Science education by connecting
  several subject areas in a principled manner. The first release of
  \fsl{} (V1), as discussed in this paper, was built through a
  relatively standard methodology involving common steps for
  expectations, reuse, conceptualization, formalization, and
  validation. We leveraged \genai{} to support ontology engineering
  (discovery, classification, linkage, completion, and transformation).}


\begin{document}

\section{Introduction}

This work presents a significant effort towards building an ontology
for the foundations of software languages~---~\fsl{} for short. This
emerging ontology is a knowledge resource that is meant to support
Computer Science education across fields such as programming language
theory, software engineering, and software language
engineering. \fsl{} and accompanying artifacts regarding validation
etc.\ are available
online.\footnote{\url{https://github.com/softlang/fsl}} The present
paper covers the first release of \fsl{} (V1) and fully presents its
development.  In the remainder of the introduction, we discuss the broader
context of \fsl, i.e., software language engineering (\sle), we
motivate the need for an ontology for the foundations of software
languages (\fsl), we demonstrate the ontology at work (`in a
nutshell'), we summarize contributions and methodology (based on
ontology engineering), and we conclude with the roadmap of the paper.


\subsection{Background~---~Software Language Engineering}

The notion of `software language'~\cite{FavreGLW09,SoftLangBook18}
integrates more specialized views on different types of `computer
languages', as they have developed in different fields of computer
science, for example, i) programming languages and domain-specific
languages in the context of programming language theory and language
implementation including compiler construction; ii) modeling languages
in the context model-driven engineering; iii) query languages in the
context of databases or semantic data. Basically, every artificial
language used in the software lifecycle and in software engineering is
a software language. The encompassing notion of `software language
engineering' assumes that software languages themselves require
engineering, including lifecycle support, and that one needs to
consider them often in the context of different technological
spaces~\cite{KurtevBezivinAksit2002TechnologicalSpaces}. The \sle{}
community, as perhaps best identifiable through its conference
series,\footnote{\url{https://dblp.org/db/conf/sle/}} over the last 20
years, has made countless contributions on top of the aforementioned
`assumption'. Some important developments concern language
workbenches, safer metaprogramming approaches, bridges between
technological spaces, support for advanced (e.g., bidirectional)
transformation scenarios.


\begin{figure}[!htbp]
\noindent
\begin{center}
  \includegraphics[width=0.95\textwidth]{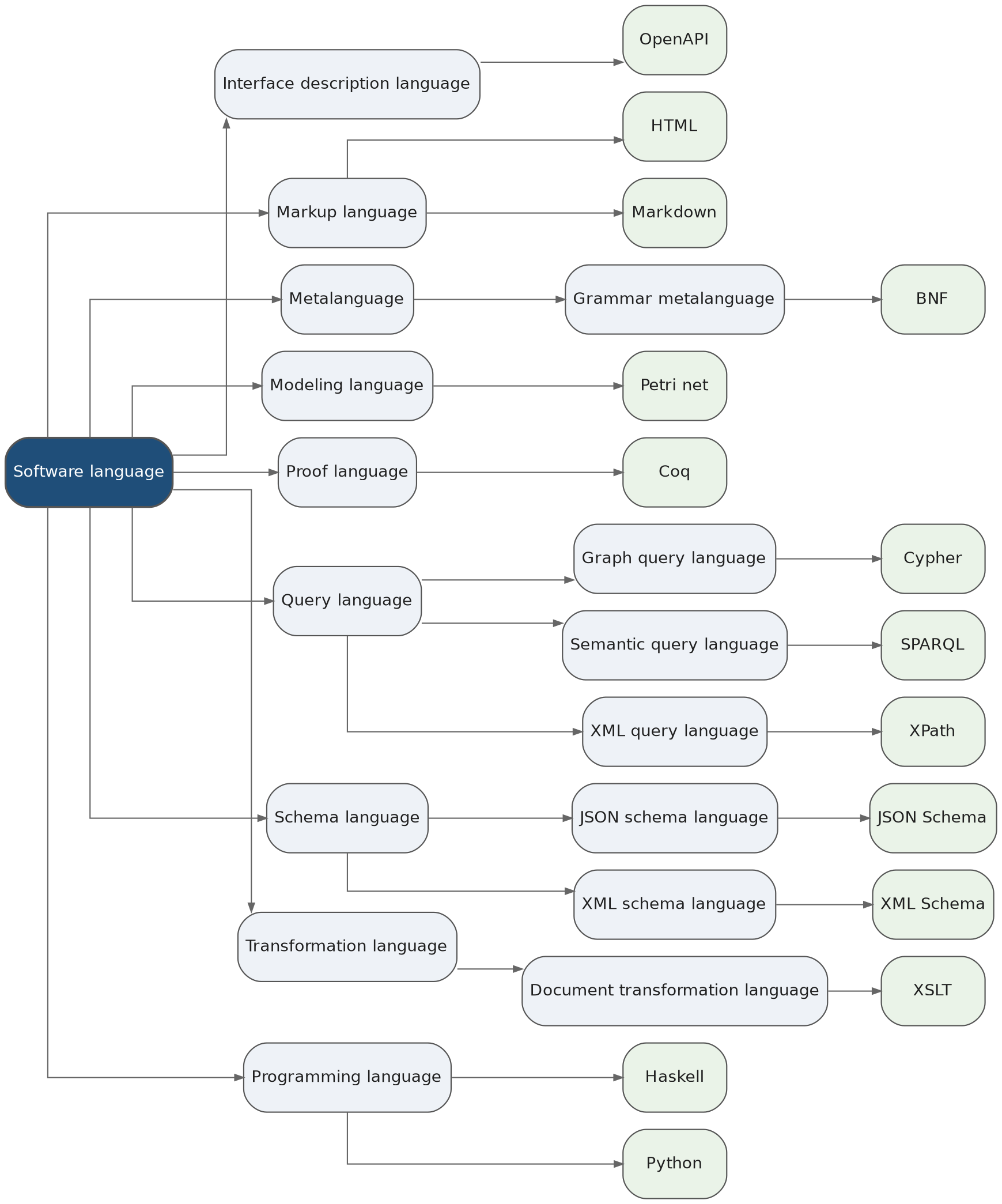}
\end{center}
\caption{Selected software language categories in \fsl}
\label{F:software_languages}
\end{figure}




\subsection{In Need of an Ontology for the Foundations of Software Languages}

As a field matures and grows, it is natural to see the need for robust
knowledge resources. \sle{} is no different in this respect, and such resources are
perhaps even more important for \sle, as it is an interdisciplinary
and knowledge-integrating field. In some of the sub-areas of \sle,
taxonomy- or ontology-like contributions have been developed. For
example, programming language
classification~\cite{BabenkoRY75,DoyleS87} is an established effort.
However, there have been no successful efforts to provide a relatively
comprehensive ontological foundation for \sle.

We will discuss all available `pre-ontological' material in some
detail, when we discuss the aspects of ontology reuse in
Sec.~\ref{S:reuse}. A few highlights are given here.

A noteworthy (and somewhat ongoing) effort has been SLEBOK (The \sle{}
Body of Knowledge) which should feature ``\emph{artifacts,
  definitions, methods, techniques, best practices, open challenges,
  case studies, teaching material, and other components that will
  afterwards help students, researchers, teachers, and practitioners
  to learn from, to better leverage, to better contribute to, and to
  better disseminate the intellectual contributions and practical
  tools and techniques coming from the SLE field}''
~\cite{SLEBOK}. Some candidate pieces of such an SLEBOK have been
published~\cite{Steimann22,DSLsInRacket24,Steimann24,MontiCoreLessons25}. By
contrast, \fsl{} focuses on formal knowledge representation as opposed
to BOK-style recipes or best practices or experience reports.

A quasi-ontological approach with relevance for \sle{} is
megamodeling~\cite{BezivinJRV05,FavreLV12,BaggeZ15} where software
systems are modeled in terms of types of artifacts, technologies, and
languages involved, subject to rich relationships including
conformance, set-membership, part-whole, and structured
correspondence. Such work has not so far led to a comprehensive ontology for the foundations of software languages.

The present paper and its \fsl{} go well beyond specialized
taxonomies and megamodels.



\subsection{The FSL Ontology in a Nutshell}
\label{S:nutshell}

Let us introduce \fsl{} by a series of illustrative query scenarios
and associated result graphs, thereby essentially exercising
competency questions, as is standard practice in ontology
engineering~\cite{KeetK25}.


\begin{figure}[!htbp]
\noindent
\begin{center}
  \includegraphics[width=0.95\textwidth]{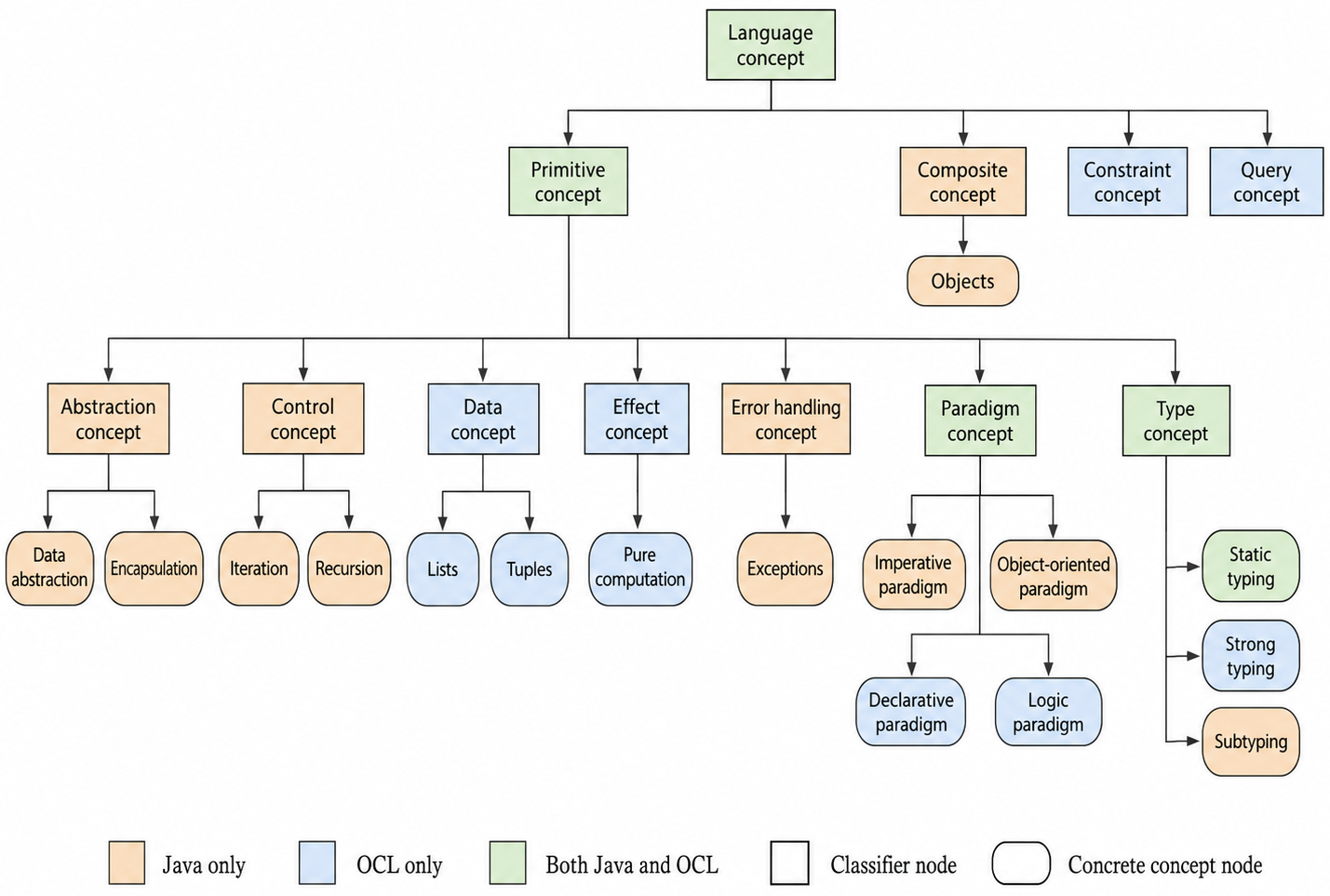}
\end{center}
\caption{Selected software language concepts in \fsl}
\label{F:language_concepts}
\end{figure}


In our first scenario, we expect to be able to query different classes
(types) of software languages with the leaves of the shown
classification hierarchy corresponding to actual
languages. Fig.~\ref{F:software_languages} shows the corresponding
query result while including only some categories and individuals for
the sake of presentability.

In our second scenario, we expect to be able to query languages in a
manner that we can retrieve the language concepts for the languages. A
simplified expectation would be here to query languages by `paradigm'.
Fig.~\ref{F:language_concepts} shows a more refined approach in
\fsl~---~while paradigmatic concepts appear (see imperative and OO
paradigm as well as declarative and logic paradigm), there are many
more refined language concepts. For example, there are concepts
related to abstraction, control, and typing. In the figure, we focus
on Java as a programming language versus OCL as a non-programming
language. We can observe what concepts are specific to either language
versus what concepts are shared. Only relatively abstract concepts are
shared; see, for example, `paradigm concept'.


\begin{figure}[!htbp]
\noindent
\begin{center}
  \includegraphics[width=0.65\textwidth]{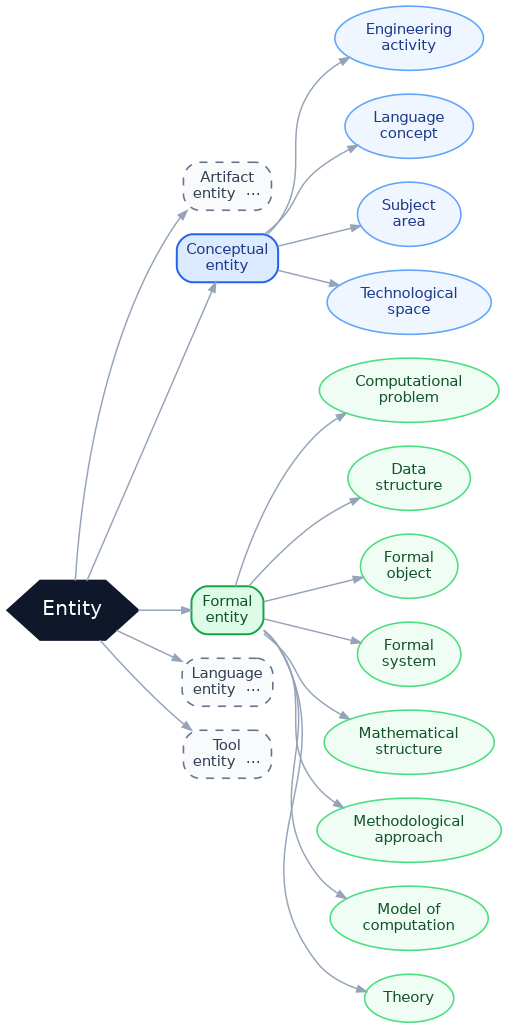}
\end{center}
\caption{\fsl{} entity types~---~top-level view}
\label{F:entity_types}
\end{figure}


\begin{figure}[!htbp]
\noindent
\begin{center}
  \includegraphics[width=0.95\textwidth]{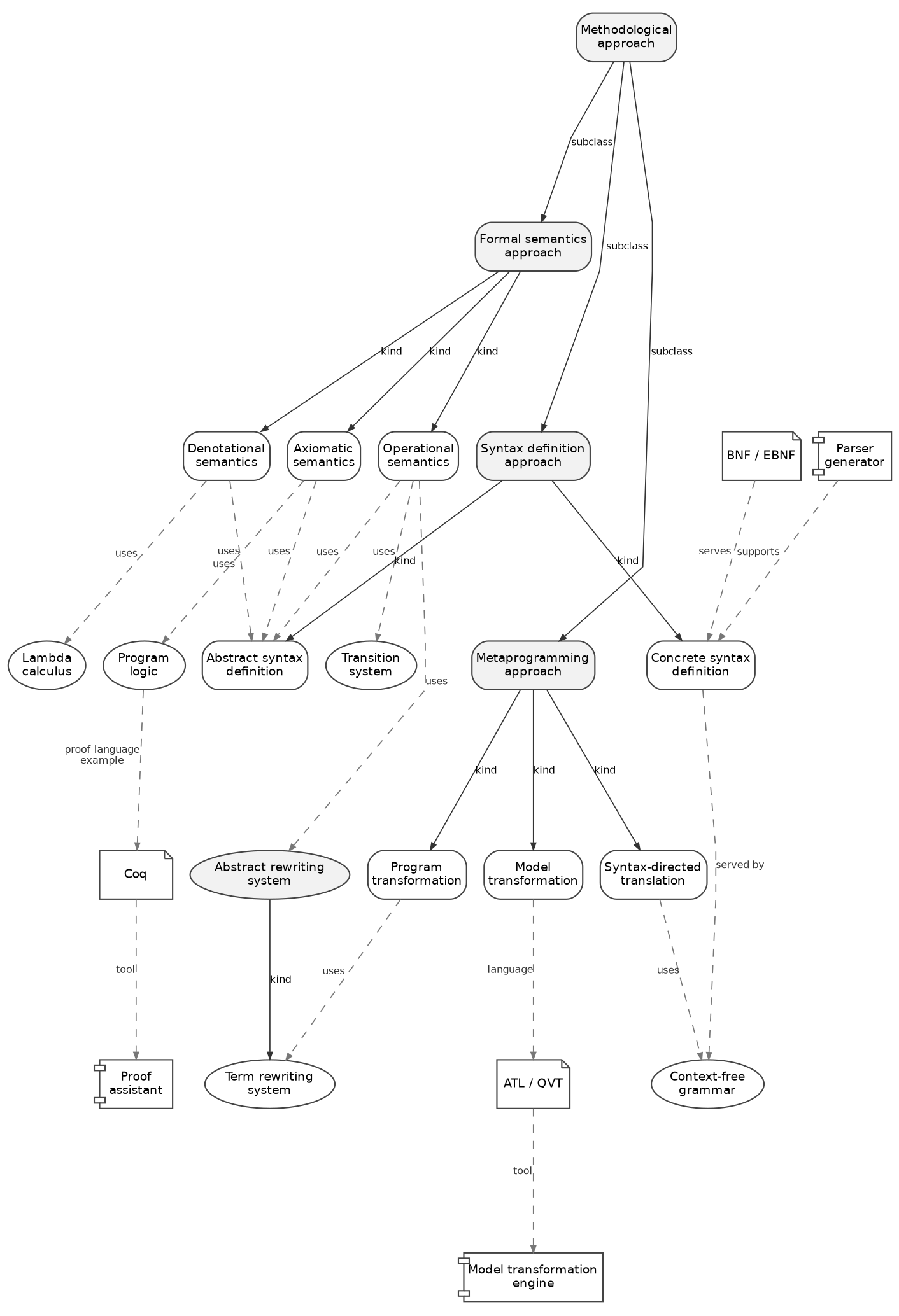}
\end{center}
\caption{Selected formal entities in \fsl{} and their associations among themselves
  and with languages and tools}
\label{F:formal_entities}
\end{figure}



Fig.~\ref{F:entity_types} is less about a competency question or an
expectation. This figure simply summarizes the top-level class
hierarchy of \fsl. The language taxonomy is intentionally left out, as
we saw already part of it in Fig.~\ref{F:software_languages}. The tool
taxonomy is left out for now; we will encounter some types of tools
below. The most general subclasses of `Conceptual entity' and `Formal
entity' are included. Let us explain a few nodes. `Formal entity' is
root to (Hilbert-style~\cite{HilbertBernays1934,HilbertBernays1939})
formal systems (such as lambda and process calculi, formal grammars,
etc.), methodological approaches (such as metaprogramming, abstract
syntax definition, and denotational semantics), and yet other formally
defined entities. `Conceptual entity' is root to different types of
semantic annotations including language concepts (exercised in
Fig.~\ref{F:language_concepts}) and technological spaces explored
separately below.

In our next, the third scenario, we expect to be able to observe how
formal entities (such as formal systems and methodological approaches)
engage in relationships among themselves and with languages and tools;
see Fig.~\ref{F:formal_entities}. For instance, we can observe that
denotational semantics uses lambda calculus and abstract syntax
definition; also, model transformation is supported by the languages
ATL and QVT and by the tool category of model transformation engines.


\begin{figure}[!htbp]
\noindent
\begin{center}
  \includegraphics[width=0.9\textwidth]{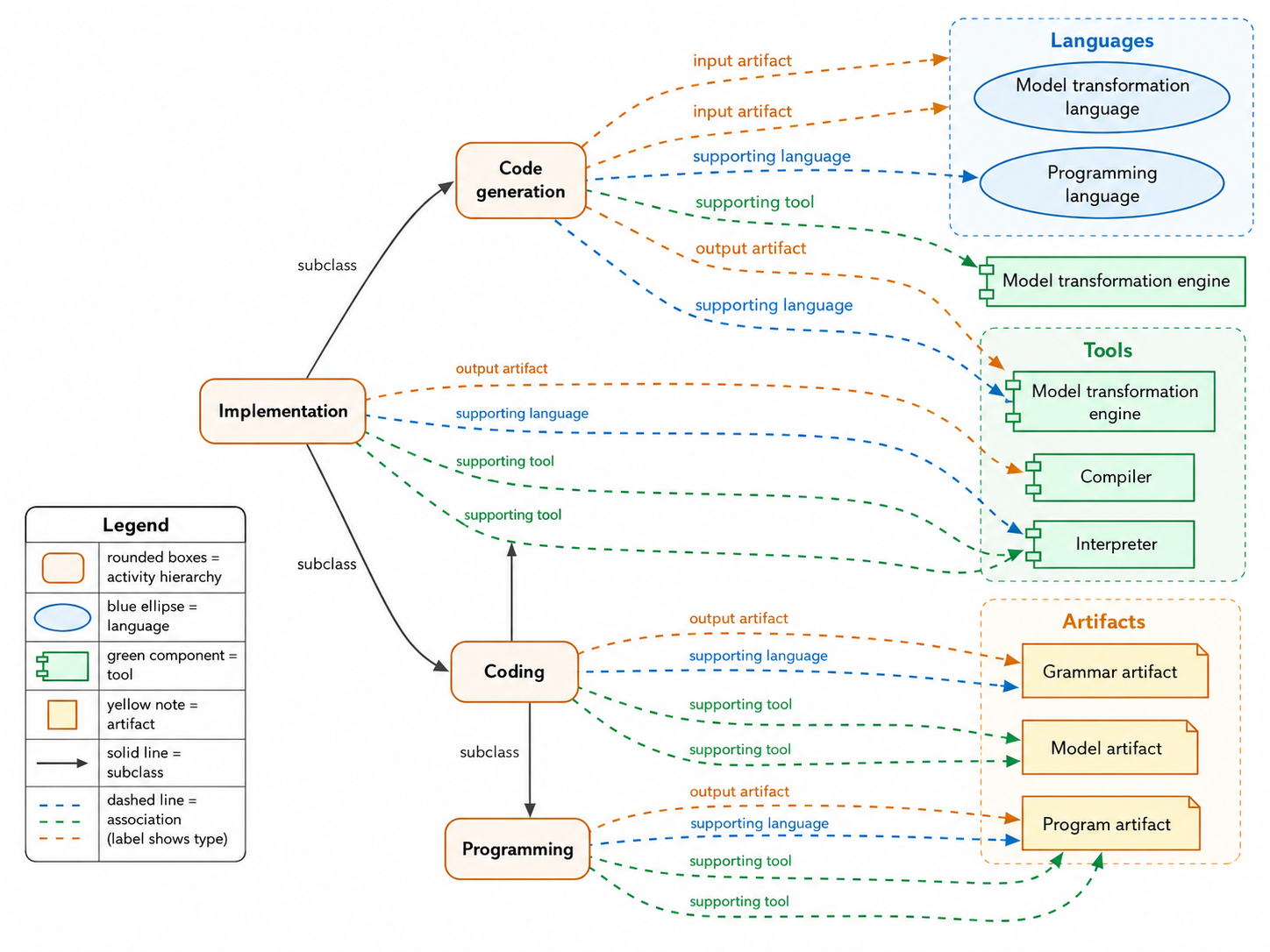}
\end{center}
\caption{Selected SE activities in \fsl (all related to implemented)
  and their associations with languages, tools, and artifacts}
\label{F:implementation_subtree}
\end{figure}


In our next, the fourth scenario, we look into software engineering
activities, while focusing here on implementation-related activities
(as opposed to, for example, requirements, design, and deployment)
and we expect to observe what kinds of languages, tools, and artifacts
are relevant here; see Fig.~\ref{F:implementation_subtree}. For
example, code generation may be supported by a model transformation
language and a model transformation engine.


\begin{sidewaysfigure}[!htbp]
  \includegraphics[width=.85\textheight]{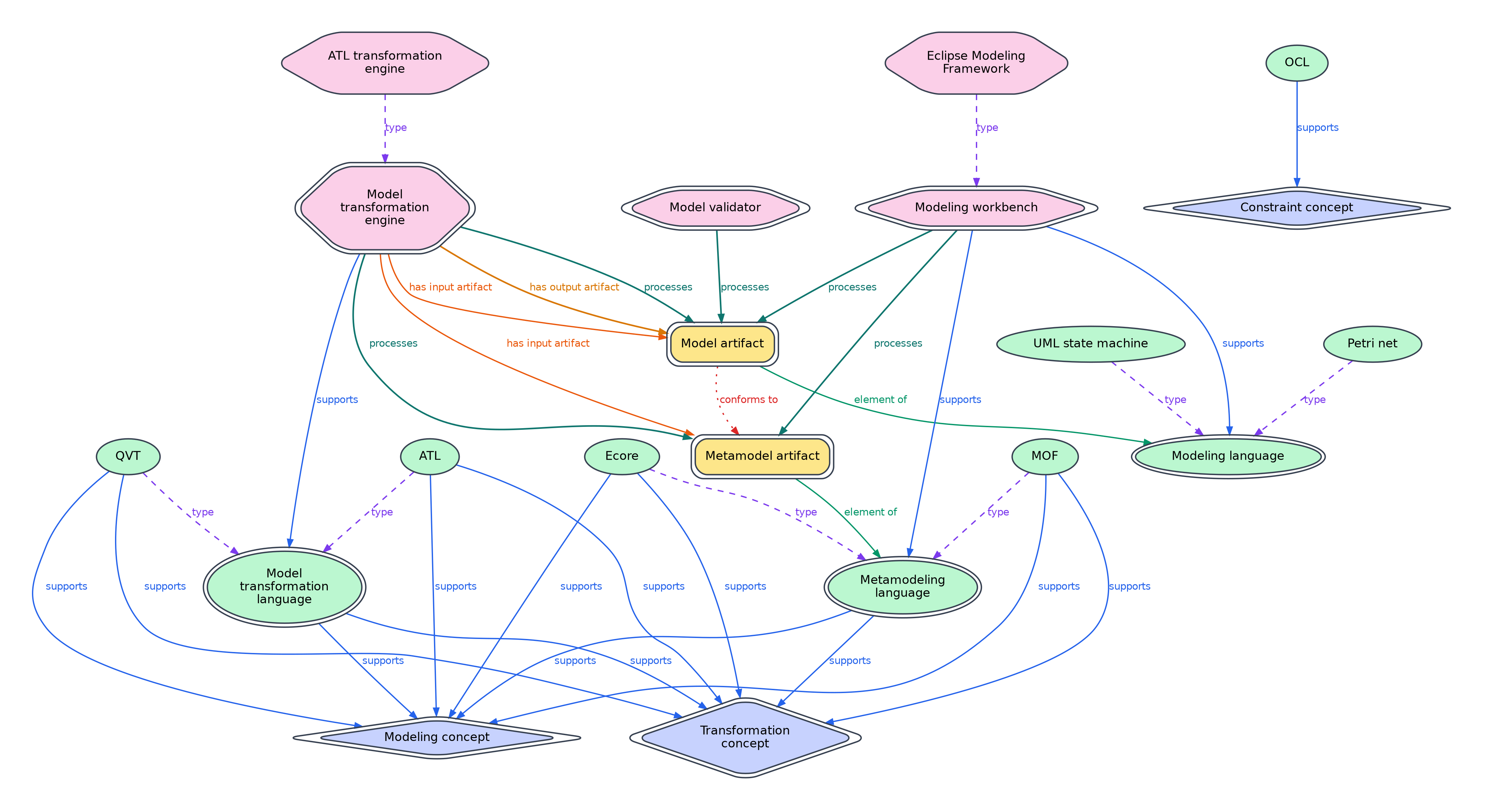}
  \caption{The technological space of Modelware in FSL}
  \label{F:modelware}
\end{sidewaysfigure}


In our last, the fifth scenario, we expect to understand the typical
ingredients of a technological space. That is, we query for all (types
of) languages, tools, artifacts, formal, and conceptual entities as
they are associated with the space; we also expect to observe the
relationships between all these associated
entities. Fig.~\ref{F:modelware} shows the resulting model for
`Modelware'~\cite{Bezivin06}.




\subsection{Contributions and Methodology}

We provide the first relatively comprehensive effort to organize the
field of software languages including their engineering in an
emerging ontology derived by a relatively formal and standard ontology
engineering process. All previous efforts (see Sec.~\ref{S:slepreonto}
specifically) are either focused on rather specific aspects (e.g.,
programming language classification) or they do not leverage ontology
engineering (like in our own work on 101companies~\cite{FavreLSV12}).

Our methodology is relatively standardized in terms of executing
ontology engineering with the following specifics. Obviously, the
methodology is specialized towards software languages and their
engineering. In this context, we invoke crucial notions such as
technological spaces, language concepts, and software engineering
activities. Much in alignment with current trends in ontology engineering,
we leverage \genai{} to semi-automatically perform some steps of ontology
engineering. We provide explained traces of \genai{} usage.


\subsection{Roadmap of the paper}

Sec.~\ref{S:mm} develops the `Materials and Methods' of our
research. At its heart, the section outlines nine phases of bringing
\fsl{} into existence. There is also a part on the use of
\genai{}. Sec.~\ref{S:results} presents quantitative and qualitative
results based on the execution of the phases of our
methodology. Sec.~\ref{S:discuss} summarizes our findings and presents
future work. Sec.~\ref{S:concl} concludes the paper.

\section{Materials and Methods}
\label{S:mm}

Sec.~\ref{S:oe} reviews the method of \emph{ontology engineering} and explains how it is applied in the methodology for \fsl{}.  Sec.~\ref{S:reuse} focuses on an important aspect of ontology engineering, i.e., \emph{reuse}, to clarify what the knowledge resources underlying \fsl{} are and to what extent existing foundational ontologies have been leveraged for \fsl. Importantly, Sec.~\ref{S:phases} describes the \emph{phases} of developing \fsl{} in terms of the (non-) goals for each stage. Finally, Sec.~\ref{S:genai} explains the use of \genai{} in this research and publication effort.



\subsection{Ontology Engineering}
\label{S:oe}

\fsl~---~the key result of the paper~---~has been created by means of an ontology engineering methodology. Accordingly, we rehash the field of ontology engineering below. We begin by recalling the basics in terms of scholarly work that paved the way. Afterwards, we summarize the general process for ontology engineering. As our ontology engineering effort employs \genai{} in an essential manner, we also discuss related work on this aspect. Finally, as we are particularly interested in validating \fsl{} for its first release, we also discuss related work on this aspect.


\subsubsection{Ontology Engineering~---~Basics}

Ontology engineering is a mature field, perhaps starting in the early 1990ies with the notion of portable, reusable ontology specifications~\cite{Gruber1993Portable}, with the clarification of foundational principles, methods, and applications~\cite{UscholdGruninger1996Principles}, with moving methodology from `ontological art' to engineering~\cite{FernandezLopez1997Methontology}, also leading eventually (towards the end of the 1990ies) to a state-of-the-art in ontology engineering~\cite{GomezPerez1999StateOfArt,StaabEtAl2001KnowledgeProcesses} and to the availability of a clear guideline for building ontologies~\cite{NoyMcGuinness2001Ontology101}.

It is instructive to observe the characteristics of more recent ontology engineering work. NeOn with its scenario-based, reuse-heavy methodology is quite established~\cite{SuarezFigueroaEtAl2015NeOn}. Challenges regarding ontology languages (e.g., the path from SKOS to OWL) and effective reuse (for sizable ontologies) are well pronounced~\cite{Tudorache2020CurrentState}. The (necessary) alignment of the industrial lifecycle aligned with Semantic Web practice (LOT)~\cite{PovedaVillalonEtAl2022LOT} has received much interest; the need for an agile and collaborative methodology for organizations as well~\cite{SpoladoreEtAl2023AgiSCOnt}. Also, the combination of ontologies and knowledge graphs has been a recent concern~\cite{PernischEtAl2025LOT4KG}. LOT~\cite{PovedaVillalonEtAl2022LOT} and NeOn~\cite{SuarezFigueroaEtAl2015NeOn} can serve as the backbone of any current methodology (like ours).


\subsubsection{Ontology Engineering~---~the Recommended Process}
\label{S:process}

A sensible current process, distilled from \cite{FernandezLopez1997Methontology,NoyMcGuinness2001Ontology101,StaabEtAl2001KnowledgeProcesses,SuarezFigueroaEtAl2015NeOn,PovedaVillalonEtAl2022LOT,SpoladoreEtAl2023AgiSCOnt,Tudorache2020CurrentState}, is:

\begin{description}

\item[Expectation] Specify purpose, scope, users, and competency questions.

\item[Reuse] Survey and reuse existing ontologies, patterns, vocabularies, and data models first.

\item[Conceptualization] Conceptualize with domain experts, iteratively and collaboratively.

\item[Formalization] Formalize (`encode') in OWL et al.\ and document decisions.

\item[Validation] Validate continuously: logical consistency, competency-question coverage, data-shape validation, and quality checks.

\item[Release] Publish, version, and maintain the ontology as part of a broader KG/data lifecycle, also attracting and incorporating feedback from the relevant community.

\end{description}

We follow this process in a refined manner; we also acknowledge that \fsl{} is a very early stage of development. The expectation has been set in the introduction (specifically Sec.~\ref{S:nutshell}) with a number of scenarios (competency questions). Reuse is considered below (Sec.~\ref{S:reuse}). Conceptualization, formalization, and validation are covered by our phases presented below (Sec.~\ref{S:phases}). Domain experts, up to this point, were the author and \genai. \fsl{} has been released on GitHub; a more advanced release process for new versions is beyond the scope of this paper, but see the future work discussion on CI/CD (Sec.~\ref{S:cicd}).


\subsubsection{Ontology Engineering~---~Use of AI (NLP, \genai)}

Such use is, of course, established by now. (We focus on \genai{} usage in what follows.) For example, ontology learning with \genai{} is considered in a various places~\cite{BabaeiGiglouEtAl2023LLMs4OL,ShimizuHitzler2025Accelerating,GarijoEtAl2025LLMTasks,ValCalvoEtAl2025OntoGenix, LiPovedaGarijo2025SLR,KamparsEtAl2025Collaborative}; more specifically, a survey of LLM tasks in ontology engineering has been compiled~\cite{GarijoEtAl2025LLMTasks} and the problem LLM-assisted ontology engineering from datasets has been researched~\cite{ValCalvoEtAl2025OntoGenix}; see also the systematic literature review on LLMs for OE~\cite{LiPovedaGarijo2025SLR}. Our work is a response to the programmatic call to accelerate work on knowledge graphs and ontology engineering with LLMs~\cite{ShimizuHitzler2025Accelerating}. Our methodology is `hybrid' in how human–\genai{} collaborative ontology design with expert validation is performed~\cite{KamparsEtAl2025Collaborative}; in our approach, the initiative remains fully with the human author; \genai{} helps with productivity.


\subsubsection{Ontology Engineering~---~Validation}

An overall assumption in validation is that structural, functional, and usability-profiling measures need to be `tested'~\cite{GangemiEtAl2006EvaluationValidation}. There is also the notion of ontology quality assessment~\cite{WilsonEtAl2023QualityAssessment}. There is actually tooling explicitly for ontology evaluation or validation~\cite{HammoudaEtAl2024MoOnEv} such that the many aspects are modularly covered. Validation may also rely on systematic use of constraints~\cite{FerrantiEtAl2024WikidataSHACL} (here: a combination of SHACL + SPARQL + Wikidata constraints) and there is actually a tricky interplay between validation and inference~\cite{RobaldoBatsakis2024SHACLTime} (in the context of temporal ontology constraints). Of course, LLMs are also used for validation, for example, by means of translating textual requirements into executable validation queries~\cite{TufekEtAl2024SemanticArtifacts}. Validation in terms of effective use of Wikipedia (by means of `browsing' also)~\cite{YuThomTam2007WikipediaEvaluation} is also heavily exercised in our methodology.



\subsection{Reuse-related aspects}
\label{S:reuse}

Reuse of knowledge resources (preferably, ontologies) is a key requirement in ontology engineering, which we want to address here upfront. We look at two directions for possible reuse: a) upper ontologies that may provide a foundation for the more domain-specific ontology \fsl{} we are aiming at; b) any knowledge structures in the immediate field of software languages that could suggest entities and relationships for \fsl{}.


\subsubsection{Upper ontologies with reuse potential}

\fsl{} reuses some standard ontologies, for example, FOAF\footnote{\url{https://xmlns.com/foaf/spec/}} and SKOS\footnote{\url{https://www.w3.org/2004/02/skos/}}, but the interesting question is whether \fsl{} could reuse, in a meaningful manner, an upper ontology. That question is a common one in ontology development~\cite{PartridgeEtAl2020}. Let us discuss common options.

The Basic Formal Ontology (BFO)\footnote{\url{https://github.com/bfo-ontology/BFO/wiki}} is an upper ontology often reused across sciences; it focuses on space and time. Currently, space is not relevant for \fsl. Time would be relevant for \fsl, if we were to cover, for example, processes properly, for example, in software development. The General Formal Ontology (GFO)\footnote{\url{https://www.onto-med.de/ontologies/gfo}} is somewhat similar to BFO, but it attempts to include several aspects of modern philosophy. GFO includes functions (in the teleological sense, i.e., relating to purpose); they could be useful in coordinating purpose-related dimensions in \fsl. 

The upper ontology of Business Objects Reference Ontology (BORO) \cite{Partridge1996}~---~not available in OWL~---~can be used, for example, to model legacy software systems in the context of software re-engineering. \fsl{} does not go into any depth of software architecture. The CIDOC Conceptual Reference Model\footnote{\url{https://cidoc-crm.org/}} is an ontology for the cultural heritage domain with a subset, CRM Core, which is an upper ontology for space-time, material and immaterial things, whereas \fsl{} is mostly about immaterial things. The COmmon Semantic MOdel (COSMO) ontology\footnote{\url{https://micra.com/COSMO/}}$^,$\footnote{\url{https://github.com/mauna-ai/COSMO.owl}} is an upper ontology for semantic interoperability (i.e., an ontology to enable software systems to exchange data with unambiguous, shared meaning). Because of its focus on contemporary English (as opposed to the highly specific nature of software languages and their foundations) and also because of its huge size, it is not straightforward to extract a fragment with relevance for \fsl{} from it.

The Descriptive Ontology for Linguistic and Cognitive Engineering (DOLCE)\footnote{\url{https://www.loa.istc.cnr.it/index.php/dolce/}} (also its OWL version DOLCE-Ultralite\footnote{\url{http://www.ontologydesignpatterns.org/ont/dul/DUL.owl}}) is an upper ontology with classes for `objects' (any physical, social, or mental object to participate in processes and to be spatiotemporally located), `events', `processes', `qualities' (such as shape), `abstracts' (i.e., non-spatiotemporal entities) with relationships for participation (of objects in processes), inherence (for qualities to inhere in entities), parthood (i.e., mereological relations), and dependence (i.e., qualities depending on their bearers). Except for some aspects, such as mereology, we cannot easily judge whether the overall class structure of DOLCE provides a helpful foundation for \fsl. However, DOLCE-Ultralite has paved the way for Ontology Design Patterns\footnote{\url{http://ontologydesignpatterns.org/index.php/Ontology_Design_Patterns_._org_(ODP)}}$^,$\footnote{\url{https://dblp.org/db/conf/wop/index.html}}$^.$\footnote{\url{https://github.com/odpa}}~\cite{ODP2009}, which are of use for \fsl, for example, Classification, ConceptTerms, Constituency, Description, Information realization, Partition, PartOf, Tagging, Topic. We use such patterns without any formal import. As a descendant of the BFO and DOLCE line of work, there is now also gUFO: a gentle foundational ontology for semantic web knowledge graphs~\cite{gUFO}. We may align \fsl{} with gUFO at some point.


\subsubsection{Pre-ontological knowledge in the software languages field}
\label{S:slepreonto}

Textbooks on programming languages and paradigms (such as~\cite{Stoy77,Mosses92,Gunter92,Tennent94,SlonnegerKurtz95,Sethi96,Strachey00,TAPL,TAPL2,NielsonNH04,Krishnamurthi07,PLessentials08,Scott09,Sestoft12,Sebesta12}) are rich knowledge resources along multiple dimensions such as syntax, static semantics (well-formedness and typing), dynamic semantics (e.g., based on operational semantics), pragmatics, paradigms, language concepts and mechanisms. We make sure that \fsl{} covers such dimensions. We do not aim at the most detailed coverage of these programming language-related dimensions, but we instead aim to generalize across other types of software languages. Few of these textbooks are in a conceptualized format. One could call out Mosses' action semantics~\cite{Mosses92}, as it features a highly structured system of (tiny) executable language definition modulets.

Research on domain-specific languages (DSLs) (or more specifically, domain-specific modeling languages (DSMLs)) and the corresponding research areas of generative programming, and software language engineering~\cite{CzarneckiE00,Wile03,Wile04,MernikHS05,GrayFCKMT08,CehCKM11,dslbook13,GlobalizedDSLE14,SchaussEtAl17,SoftLangBook18,Dsl23} expose the lifecycle of software languages and many different ecosystems for supporting the lifecycle, thereby also suggesting some aspects of \fsl. Some of this work is so highly structured that it directly suggests entities and relationships of \fsl. For instance, L\"ammel, Schauss et al.'s work~\cite{LaemmelLSV14,SchaussEtAl17} on a chrestomathy of DSL implementations systematically organizes features of language implementation based on metaprogramming systems and also adds semantic labels for different classes, for example, technology, programming concepts, and involved languages.

Classification hierarchies (taxonomies) or slightly richer quasi-ontological structures have been proposed for a number of types of software languages: programming languages~\cite{BabenkoRY75,DoyleS87}, visual languages~\cite{MarriottM97}, model transformation languages (and approaches and tools)~\cite{MensG06,CzarneckiH06,TamuraC10,GomesBA14}, business rules modeling languages~\cite{SkalnaG12}, software architecture description languages~\cite{MedvidovicT00}, and possibly others.  Domain-specific languages~---~as a larger container of software languages~---~were also categorized by means of a mapping study~\cite{KosarBM16}.  Even the entirety of software languages (or computer languages per Wikipedia terminology) were categorized through a mining and cleaning-based approach aiming at alignment with Wikipedia categories~\cite{LaemmelMV13}. Rather than aiming at the classification of languages, there are additional categorization efforts covering some aspects of programming (programming languages) or software engineering, for example, a classification of typing approaches in programming languages~\cite{CardelliW85}, a taxonomy for software re-/reverse engineering~\cite{ChikofskyC90}, and a chrestomathy of programming techniques and technologies relying heavily on feature modeling and other means of semantic data enrichment~\cite{FavreLSV12}.

A quasi-ontological approach in software engineering is megamodeling~\cite{BezivinJRV05,FavreNguyen2004MegamodelEvolution,FavreLaemmelVaranovich2012LinguisticArchitecture,LaemmelV14,BaggeZ15,HaertelHHLV17,Laemmel21} where software systems are modeled in terms of types of artifacts, technologies, and languages involved, subject to rich relationships including conformance, set-membership, part-whole, and structured correspondence. These megamodels provide ontological insights into particular patterns of technology usage or system development, for example, 
parsing~\cite{ZaytsevB14}, model-driven engineering~\cite{RoccoRHILP20}, or software transformations (coupled transformations, specifically)~\cite{Laemmel16}. Megamodels also relate to the notion of technological spaces~\cite{KurtevBezivinAksit2002TechnologicalSpaces}, which can be seen as an ontological pattern relating primary and secondary artifact types, some notion of schema or metamodel that supports conformance, and some forms of languages for querying, transformation, and others. Typical technological spaces are grammarware~\cite{KlintLaemmelVerhoef2005Grammarware}, model-driven engineering~\cite{Bezivin2006MDETechnicalSpace}, SQLware, XMLware, JSONware, and Ontoware. We incorporate some central aspects of megamodeling and technological spaces into (the first release of) \fsl.

Software language engineering is software engineering for languages. Thus, the `body of knowledge' on software engineering applies~\cite{BourqueDAMT99,DupuisB00,Bourque20}. There are also other bodies of knowledge in neighboring territories, most notably in model-based engineering~\cite{BurguenoCFKLMPP19}. We can draw from these BOKs for software engineering-related entity types in \fsl.



\subsection{Phases of \fsl{} Development}
\label{S:phases}

\newcommand{\phaseI}{Phase 1~---~Taxonomy Initialization}
\newcommand{\phaseII}{Phase 2~---~Category Discovery}
\newcommand{\phaseIII}{Phase 3~---~Taxonomy Enrichment}
\newcommand{\phaseIV}{Phase 4~---~Property Discovery}
\newcommand{\phaseV}{Phase 5~---~Technological Space Coverage}
\newcommand{\phaseVI}{Phase 6~---~Modularization}
\newcommand{\phaseVII}{Phase 7~---~SL Concept Coverage}
\newcommand{\phaseVIII}{Phase 8~---~SE Activity Coverage}
\newcommand{\phaseIX}{Phase 9~---~Review}

The overall objective was to arrive at a `MVP'-like (minimum viable product) first release of \fsl, to be able to publish this first release, to enable its use and review, to publish the underlying ontology engineering (\ontoeng) effort in a scholarly manner, and thereby to lay the foundation of continuing the work at a larger scale with more domain experts and contributors involved. To this end, we went through the phases as summarized in Table~\ref{T:phases}.

\begin{table}[ht]
  \small
  \noindent
  \begin{center}
  \begin{tabular}{|l|}\hline\hline
    \phaseI\\\hline
    \phaseII\\\hline
    \phaseIII\\\hline
    \phaseIV\\\hline
    \phaseV\\\hline
    \phaseVI\\\hline
    \phaseVII\\\hline
    \phaseVIII\\\hline
    \phaseIX\\\hline\hline
  \end{tabular}
  \end{center}
  \caption{The phases of ontology engineering for V1 of \fsl.}
  \label{T:phases}
\end{table}

Here is a summary per phase:

\begin{description}
\item[\phaseI] (Sec.~\ref{S:phase1}) This is the trivial starting point. Using a manually designed small seed set of \sle{} formalisms, we begin with a simple taxonomy. We start exercising an overarching \ontoeng{} principle, i.e., to connect all \fsl{} resources, as much as possible, to knowledge resources. We assume that the seed set will effectively serve as a focus group for validation in any of the subsequent phases.
\item[\phaseII] (Sec.~\ref{S:phase2}) We deliberately do not distinguish categories versus individuals in \emph{Phase 1}. Instead, we establish this distinction separately in this phase so that we can perform a small test regarding the use of \genai{} for individual/category distinction (i.e., classification).  We have certain expectations regarding the categories to be discovered. For example, we have multiple closely related calculi in the seed set. 
\item[\phaseIII] (Sec.~\ref{S:phase3}) With some categories in place, we want to make the classification more meaningful by aiming at a very limited completion effort focusing on categories and individuals in the immediate `neighborhood' of the existing entities. As a result, the new entities are in the scope of our expertise, which simplifies the task of validating the use of \genai{} for completion.
\item[\phaseIV] (Sec.~\ref{S:phase4}) We begin to introduce properties for \fsl{} entities such as `uses' and `isSpecifiedBy'. We leverage the fact that \fsl{} features different types of entities, in particular: more calculus-like entities (e.g., lambda calculus) versus more approach-like entities (e.g., denotational semantics). Property discovery is continued through phases~5, 7, 8.
\item[\phaseV] (Sec.~\ref{S:phase5}) Based on our understanding of technological spaces as an important quasi-ontological concept in the \sle{} context, we integrate such spaces into the emerging ontology. That is, we agree on a number of established technological spaces as well as relationships between them and languages, tools, artifacts, and formalisms. Tools and artifacts are introduced at this point, subject to new root classes of the ontology.
\item[\phaseVI] (Sec.~\ref{S:phase6}) The emerging ontology eventually requires modularization to remain manageable and understandable. Modularization can centralize the tbox and group entities based on the top-level entity type. That is, on top of the tbox, there are aboxes for languages, tools, formalisms (and such), and more conceptual entities. Naming and details of modular decomposition can evolve over the remaining phases.
\item[\phaseVII] (Sec.~\ref{S:phase7}) An important partnering subject area of \sle, also known for rich knowledge resources, is `programming language theory' and `programming paradigms' or more broadly `programming language concepts'. At this stage, we aim at capturing programming language concepts in \fsl, associating programming language entities with them and generalizing the hierarchy of programming language concepts to become meaningful for all software languages.
\item[\phaseVIII] (Sec.~\ref{S:phase8}) Likewise, \sle{} is software engineering for languages; also, languages serve software engineering. Thus, we shall also determine a more SE-centered sub-taxonomy to serve as a foundation for linking the rest of \fsl{} to software engineering. We suggest that SE shall be approached in terms of SE activities (such as requirements, design, implementation, etc.), with refinement into more specific activities (such as code generation under implementation). In this manner, those activities can also be associated with languages, tools, artifacts, etc.\ or, in fact, types thereof.
\item[\phaseIX] (Sec.~\ref{S:phase9}) Up to this point, we primarily pushed for ontology completion with validation and normalization as subordinate goals. During this phase, we review the emerging \fsl{} ontology in depth. For example: What (SHACL) shapes should be devised? How could classes be defined definitionally or what disjointness should be claimed? Also, we devise a set of (SPARQL) queries to summarize the emerging knowledge graph at the meta-level, thereby helping us to arrive at a regular and explainable structure.
\end{description}

We now describe the phases in detail. For each phase, we present goals and non-goals. The overall phase-like process, as summarized above, was designed up front as a methodology, but some details (a few of the more subordinate goals or non-goals) in the following description only emerged during execution of the process.


\newcommand{\goaltla}[1]{\textsl{#1}}
\newcommand{\goalintro}[2]{\textit{#1} (\textsl{#2})}


\subsubsection{\phaseI}
\label{S:phase1}

This phase relates to the `expectation' step and (the beginning of) the `conceptualization' step in the process for ontology engineering; see Sec.~\ref{S:process}. That is, we specify our expectation by identifying a seed set of stereotypical entities of interest to form a taxonomy. We also specify prioritized subject areas of interest to be associated with the entities in the seed set. We submit to the following (non-) goals:

\begin{description}
\item[\textbf{\textit{Goals}}] \mbox{}
  \begin{description}
  \item[\goalintro{Subject Area Agreement}{SAA}] We agree on subject areas of interest (such as software language engineering or programming language theory) and prioritize them.
  \item[\goalintro{Seed-Set Agreement}{SSA}] We agree on the seed set of entities while respecting the aforementioned priorities, that is, the higher the priority of an area, the more associated entities in the seed set. \fsl, in this phase and any phase to come, must cover the seed set.
  \item[\goalintro{Knowledge Resource Linkage}{KRL}] We want all elements from the seed set to be linked to external (knowledge) resources. This could be, for example, DBpedia or Wikipedia. In this phase, we look up those resources manually to be certain.
  \end{description}
    \item[\textbf{\textit{Non-goals}}] \mbox{}
      \begin{description}
        \item[\goalintro{Individual/Category Distinction}{ICD}] At this stage, we do not aim yet at distinguishing individuals versus categories. We withhold some expectations for such a distinction until later.
\end{description}
\end{description}

This is obviously a very low bar, but still a good way to get off the ground.


\subsubsection{\phaseII}
\label{S:phase2}

We have not yet distinguished individuals versus categories (see the non-goal \goaltla{ICD} above)~---~in part to test whether automated (\genai-based) classification (category discovery) can be used in our domain context. Once we have established categories, we can expect to use them for few-shot classification subsequently, as more entities arise. We submit to the following (non-) goals:

\begin{description}
\item[\textbf{\textit{Goals}}] \mbox{}
  \begin{description}
  \item[\goalintro{Individual/Category Distinction}{ICD}] This was a non-goal before. Some of the entities in the seed set may be categories; alternatively, we may also need to introduce some categories to group entities. For instance, we expect to discover some category related to the calculi in the seed set. In the end, all seed-set entities must be categorized or serve as a category.
\end{description}
   \item[\textbf{\textit{Non-goals}}] \mbox{}
      \begin{description}
      \item[\goalintro{Subject Area Preservation}{SAP}] During process execution, we realized that it would be challenging, if not distracting, to maintain \goalintro{Subject Area Agreement}{SAA} along with enriching the taxonomy~---~also in the view of the inital priorities serving as a constraint. Thus, we allowed ourselves to continue, for now, without subject areas.
      \end{description}
    \end{description}


\subsubsection{\phaseIII}
\label{S:phase3}
         
As we have started with a fairly small seed set, there is much room for completion. For now, we make a very conservative step. That is, essentially, we look for siblings of entities and classes already at hand. This again serves as small test in \genai-based completion. We submit to the following (non-) goals:

\begin{description}
\item[\textbf{\textit{Goals}}] \mbox{}
  \begin{description}
  \item[\goalintro{Double Instantiation Enforcement}{DIE}] There must not be a class that is instantiated by not at least two individuals because this would be a sign of spurious or overfitted or insufficiently exercised classification.
  \item[\goalintro{Double Subclassing Enforcement}{DSE}] There must not be a class that is subclassed by only one class  for essentially the same reason as under \goaltla{DIE}.

  \item[\goalintro{OWL 2 Punning Enforcement}{OPE}] The categorical and the individualized levels cannot always be fully separated in ontology engineering. Consider the following example in the \fsl{} domain: `process calculus' may be an entity of interest in certain assertions, but it can also serve as a classifier for more specific process calculi. Therefore, we employ OWL 2 punning (metamodeling) in \fsl.
\end{description}
\item[\textbf{\textit{Non-goals}}] \mbox{}
  \begin{description}
      \item[\goalintro{Individual/Category Completeness}{ICC}] In this phase, we limit ourselves to discover only entities needed to address goals \goaltla{DIE} and \goaltla{DSE}. In general, the first release of \fsl{} does not aim at a reasonable complete taxonomy, but only at satisfying certain coverage criteria.
\end{description}
\end{description}


\subsubsection{\phaseIV}
\label{S:phase4}

We have focused on classification so far. In this phase, we aim at discovering relationships other than subclassing and classification on ontological entities so that we move beyond a mere taxonomy. We submit to the following (non-) goals:

\begin{description}
\item[\textbf{\textit{Goals}}] \mbox{}
  \begin{description}
\item[\goalintro{Object Property Introduction}{OPI}] Pre-ontological related work (Sec.~\ref{S:slepreonto}) suggests relationships related to, for example, `usage' (uses) and `parthood' (hasPart); also, the formalistic nature of some ontology entities shall be exercised by a relationship `isSpecifiedBy' so that one entity (e.g., a formal system) is specified by (e.g., in the sense of its semantics) another entity. (As an aside, we also need annotation and datatype properties, in addition to object properties.) Here are illustrative examples:
  {\small
\begin{verbatim}
:DenotationalSemantics :uses :AbstractSyntax .
:AttributeGrammar :hasPart :ContextFreeGrammar .
:CommunicatingSequentialProcesses :isSpecifiedBy :OperationalSemantics .
\end{verbatim}
  }
\item[\goalintro{Entity/Assertion Discovery}{EAD}] In exploring different candidate object properties and their nuances in their semantics, new individuals and categories naturally arise or they are actively discovered for populating the new properties with assertions.
\item[\goalintro{Subject Area Recovery}{SAR}] The association of \fsl{} entities with areas is now to be modelled by an 
  `hasArea' property, without any constraints regarding prioritization, but with cardinality expectations such that entities should associate with at least 1 area and with 3 areas at maximum. Those associations are mostly discoved automatically (\genai-based) from here on. Subject area associations can be automatically validated for the original seed set (Sec.~\ref{S:phase1}).
\end{description}
    \item[\textbf{\textit{Non-goals}}] \mbox{}
      \begin{description}
      \item[\goalintro{Entity/Assertion Completeness}{EAC}] This non-goal generalizes the earlier non-goal \goalintro{Individual/Category Completeness}{ICC} so that we include assertions. That is, neither do we aim at completeness for the entities (individuals and categories) in \fsl, nor do we aim at (relative) completeness for the assertions on top of the entities that are in \fsl{} (first release). We limit ourselves to satisfying certain coverage criteria such as the one of the next phase.
\end{description}
\end{description}


\subsubsection{\phaseV}
\label{S:phase5}

As we have argued in Sec.~\ref{S:slepreonto}, technological spaces provide an important ontological structure for software languages and their foundations. By adding coverage for them in \fsl, we expect to also discover important types of entities and relationships. We submit to the following (non-) goals:

\begin{description}
\item[\textbf{\textit{Goals}}] \mbox{}
  \begin{description}
\item[\goalintro{Technological Space Coverage}{TSC}] We have to introduce a new root category of technology spaces and populate it with actual spaces. Technological spaces may be associated with languages, tools, and (types of) artifacts via a `hasSpace' property.
\item[\goalintro{Megamodeling Coverage}{MMC}] The general structure of a technological space can be understood as an abstract megamodel with each individual space corresponding to a more concrete megamodel. In this manner, we expect to recover typical relationships such as `conformsTo', `elementOf', `definedBy', `transforms', and `processedBy'; we also expect to discover the essential notions of `artifact' and `tool'. We also expect to discover certain types of languages, tools, and artifacts.
\item[\goalintro{Entity/Assertion Discovery}{EAD}~---~continued] As we decide to include certain technological spaces into \fsl, we will undoubtly need to include more entities supporting these spaces and more assertions to make the new entities participate in existing properties. For example, we need to need to include additional types of languages, for example, query and transformation languages and metalanguages.
  \item[\goalintro{Knowledge Resource Linkage}{KRL}~---~continued] Linking to DB- or Wikipedia is not sufficient for such highly specialized community notions such as megamodeling and technological spaces as well as the involved key relations such as conformance. In cases like this, we shall use scholarly references to link \fsl{} entities to supporting knowledge resources.
\end{description}
    \item[\textbf{\textit{Non-goals}}] \mbox{}
      \begin{description}
      \item[\goalintro{Entity/Assertion Completeness}{EAC}~---~continued] Completeness with regard to technological spaces specifically is also not a goal, even though one could argue that there are only finitely many. However, we want to avoid getting lost in obscure space options or nuances of abstraction (e.g., Tableware versus SQLware), as this would otherwise massively drive up the size of the ontology at an too early stage. Likewise, we do not aim at full specification of all spaces in terms of the assertions conceivable.
\end{description}
\end{description}


\subsubsection{\phaseVI}
\label{S:phase6}

At this stage, \fsl{} is expected to be of a size and complexity that it helps to introduce an explicit modular structure for better scoping and usability of the emerging ontology. We submit to the following (non-) goals:

\begin{description}
\item[\textbf{\textit{Goals}}] \mbox{}
  \begin{description}
  \item[\goalintro{Modules per Entity Types}{MET}] We aim at a simple modular structure as follows. There shall be a core module~---~\emph{tbox}~---~with all the various top-level types and relationships between them. There is one module per top-level entity type, thereby grouping all assertions by subject. There is one more module that combines all these modules. All modules but \emph{tbox} are essentially aboxes.
  \item[\goalintro{Top Level Refactoring}{TLR}] During process execution, we realized that top level of the class hierachy was not clearly enough organized to enable the modular splitting. We had to refactor the class hierarchy to have a manageable number of demarcated roots.
  \end{description}
    \item[\textbf{\textit{Non-goals}}] \mbox{}
      \begin{description}
      \item[None]
\end{description}
\end{description}


\subsubsection{\phaseVII}
\label{S:phase7}

As we have argued in Sec.~\ref{S:slepreonto}, (programming) language classification provides an important ontological dimension for software languages and their foundations. We begin by discovering programming language concepts including those related to programming language paradigms. Subsequently, we aim at generalizing the hierarchy of language concepts to cover types of software languages other than programming languages. We submit to the following (non-) goals:

\begin{description}
\item[\textbf{\textit{Goals}}] \mbox{}
  \begin{description}
  \item[\goalintro{Programming Paradigm Coverage}{PPC}] The hierarchy of language concepts must include the most established programming paradigms, for example, imperative, functional, logic and object-oriented programming.
  \item[\goalintro{PL Concept Coverage}{PCC}] The hierarchy of language concepts must include branches for concepts other than paradigms, for example, concepts related to typing, abstraction, and evaluation strategy. We do not expect that the distinction of programming paradigms can dominate the concept hierarchy; we rather assume that multiple dimensions of classification go side by side.
  \item[\goalintro{Programming Language Coverage}{PLC}] We need to perform entity/assertion discovery for actual programming languages that meet the following two criteria: a) they are popular so that the concept-based elaboration for them is meaningful; b) we are familiar enough with these languages so that we can provide or validate assertions for them such that actual languages are associated with actual concepts.
  \item[\goalintro{SL Concept Coverage}{SCC}] The hierarchy of language concepts must be scaled up further to become meaningful to language categories other than programming languages. As discussed in Sec.~\ref{S:reuse}, there exist classification schemes for various language categories; they shall be modeled~---~to some extent. For instance, we expect to include concepts immediately related to querying, transformation, and metamodeling. We also need to discover assertions for them so that actual software languages associate with actual concepts.
  \end{description}

    \item[\textbf{\textit{Non-goals}}] \mbox{}
      \begin{description}
      \item[\goalintro{Entity/Assertion Completeness}{EAC}~---~continued] There are competing classification schemes for programming languages and other types of languages; there are highly detailed (if not idiosyncratic) classification schemes for languages other than programming languages. In the current cycle, we only aim to include and exercise major concepts. We defer a proper taxonomy integration effort starting from a multitude of classification schemes to another time.
\end{description}
\end{description}


\subsubsection{\phaseVIII}
\label{S:phase8}

Software languages (and \sle{} as a discipline) are highly intertwined with software engineering (SE). We can contribute to the foundations of software languages by starting a `chase' from the software lifecycle and its underlying SE activities to associate them to related languages, tools, and artifacts or, in fact, types thereof. We submit to the following (non-) goals:

\begin{description}
\item[\textbf{\textit{Goals}}] \mbox{}
  \begin{description}
  \item[\goalintro{Software Lifecycle Coverage}{SLC}] The phases of the software lifecyle (e.g., requirements, design, implementation, deployment, maintenance) provide a good top-level layer for the hierarchy of SE activities to be included into \fsl. For example, under `implementation' we may expect SE activities such as `programming' and `code generation'. 
  \item[\goalintro{Technological Space Coverage}{TSC}~---~continued] More or less specific SE activities can be naturally associated with (types of) languages, tools, and artifacts. In this manner, we also contribute to \goaltla{TSC} because the entities associated with SE activities can also be associated with spaces (e.g., via `hasSpace').
\end{description}
    \item[\textbf{\textit{Non-goals}}] \mbox{}
      \begin{description}
      \item[\goalintro{Entity/Assertion Completeness}{EAC}~---~continued] We do not aim at a comprehensive model of SE activities and the corresponding discovery of assertions towards entity types. In particular, we do not aim at covering the SE Body of Knowledge~\cite{BourqueDAMT99,DupuisB00,Bourque20} in terms of SE activities and their relationships to (types of) languages, tools, and artifacts.
\end{description}
\end{description}


\subsubsection{\phaseIX}
\label{S:phase9}

Review is a continuous activity along ontology engineering, but we concentrate some goals in the last phase of the process for the first release of the emerging \fsl{} ontology:

\begin{description}
\item[\textbf{\textit{Goals}}] \mbox{}
  \begin{description}
  \item[\goalintro{Reasoning-Based Consistency}{RBC}] The ontology must pass (w.l.o.g.) the HermiT reasoner.
    Inconsistencies could arise from the use of constraint forms such as class disjointness or equivalence.
  \item[\goalintro{Shape-Based Validation}{SBV}] The ontology must pass (w.l.o.g.) SHACL shapes for basic validation constraints on all entity types. In particular, we may insist on certain CWA-style assertions.
  \item[\goalintro{Query-Based Reporting}{QBR}] The ontology must be explained through (w.l.o.g) SPARQL queries demonstrating the various patterns in which entities associate with each other through assertions.
  \item[\goalintro{Text-Based Transformation}{TBT}] All bulk changes due to identified shortcomings during review must be properly documented as textual specifications suitable for \genai-based execution of ontology transformations.
  \end{description}
    \item[\textbf{\textit{Non-goals}}] \mbox{}
      \begin{description}
      \item[\goalintro{Release-Blocking Issues}{RBI}] In Sec.~\ref{S:evolve}, we discuss relatively obvious limitations regarding the first release of \fsl. We do not accept any of these shortcomings as release-blocking issues because it is important to get more SL(E) experts involved and therefore to expose \fsl{} as early as possible.
\end{description}
\end{description}

\subsection{Use of Generative AI}
\label{S:genai}

\newcommand{\noGenAI}{\ensuremath{-}}
\newcommand{\littleGenAI}{{\scriptsize\ensuremath{\bullet}}}
\newcommand{\moreGenAI}{{\small\ensuremath{\bullet}}}
\newcommand{\muchGenAI}{{\large\ensuremath{\bullet}}}
\newcommand{\shr}{\ensuremath{\triangleright}\,\,}


\begin{table}
{\footnotesize
  \begin{tabular}{|l|c|l|}\hline
    \textbf{Aspect} & \textbf{AI?} & \textbf{Comment}\\\hline\hline
    \textit{\textbf{Methodology}}  & \noGenAI & Follow OE best practices and SLE needs\\\hline\hline
    \textit{\textbf{Publication}} & \littleGenAI & \\\hline
    \shr Paper structure & \noGenAI & Defined by journal and methodology\\\hline
    \shr Related work & \moreGenAI & Used for search on some specific topics \\\hline
    \shr Illustrations & \littleGenAI & Some illustrations are generated with the LLM \\\hline
    \shr Other sections and parts & \noGenAI & Not counting AI-based improvement of text\\\hline\hline
    \textit{\textbf{Ontology}} & \moreGenAI & \\\hline
    \shr Entities & \moreGenAI & \\\hline
    \shr \shr Discovery & \moreGenAI & Seeded entity discovery \\\hline
    \shr \shr Classification & \moreGenAI & Few-shot classification \\\hline
    \shr \shr Linking to Wikipedia & \moreGenAI & Entity linking (LLM+Google) \\\hline\hline
    \shr Properties & \moreGenAI & \\\hline
    \shr \shr Discovery & \noGenAI & Ontology-specific properties require expert initiative \\\hline
    \shr \shr Naming & \moreGenAI & Exploration of naming options informed by prior art\\\hline
    \shr \shr Assertions & \moreGenAI & Few-shot completion \\\hline\hline
    \shr Axioms & \moreGenAI & \\\hline
    \shr \shr Domain/range axioms & \noGenAI & Part of manual discovery process for properties\\\hline
    \shr \shr Class axioms & \noGenAI & Ontology-specific axioms require expert initiative \\\hline
    \shr \shr Property axioms & \littleGenAI & Some obvious options were auto-generated \\\hline \hline
    \shr Transformations & \muchGenAI & Textual descriptions of OWL transformations \\\hline
  \end{tabular}
}
\medskip
\caption{\genai{} usage in this paper and in the underlying research (none: \noGenAI; some: \littleGenAI, \moreGenAI, \muchGenAI)}
  \label{T:genai}
\end{table}


We now summarize the usage of \genai{} in this research. We explain why \genai{} was used in this research and how this usage improved the methodology and its execution (productivity). We also clarify why \genai{} was not used in certain scopes. GPT 5.2 or 5.3 were used for all cases of \genai{} usage mentioned subsequently. Overall, \genai{} was used to improve productivity during execution of the ontology-engineering methodology, while the initiative always remained with the researcher. See Table~\ref{T:genai} for an overview; details follow:

\begin{enumerate}[label=\arabic*.]
\item \textbf{Methodology} \genai{} played no role here because the choice of methodology was relatively straightforward within the given context. We wanted a methodology that is well in line with ontology engineering best practices, i.e., with common steps for expectations, reuse, conceptualization, formalization, and validation. The methodology is original (i.e., domain-specific) insofar as technological spaces, programming concepts, and software engineering activities are important, but these types of entities are closely related to the expertise of the author and the domain of \fsl.
  \item \textbf{Publication} This paper is almost exclusively written by the author, apart from AI-based language assistance; two relatively minor cases of \genai{} usage are worth pointing out:
    \begin{enumerate}[label=\alph*)]
    \item \textbf{Related work} The author is fully aware of the literature in the \fsl{} domain, which is also evident from the citations at hand and from the fact that much of the author's contributing work dates back quite a while (culminating in the Software Languages Book~\cite{SoftLangBook18}). However, we leveraged \genai{} to better cover topics in ontology engineering (see Sec.~\ref{S:oe}) and more specifically CI/CD in that space (see future work discussion in Sec.~\ref{S:future}). We used \genai{} for search and discovery in tandem with search on Google Scholar.  
    \item \textbf{Illustrations} Some of the figures of Sec.~\ref{S:nutshell} were created via controlled \genai{} usage to improve productivity. Validation of these artifacts was straightforward. We also used chain-of-thought prompting to separate raw data extraction from visualization, thereby simplifying validation. By contrast, all tables (like those needed for quantification of results in Sec.~\ref{S:results}) are based on custom Python/SPARQL code (included in the \fsl{} GitHub project) to avoid any sort of hallucination, to be precise regarding quantification, and to enable reproducibility.
  \end{enumerate}
\item \textbf{Ontology} With the \fsl{} ontology being the primary result of this research and with the underlying ontology engineering process being the primary concern of the research methodology, we must discuss \genai{} usage from an ontology-focused perspective.
    \begin{enumerate}[label=\alph*)]
  \item \textbf{Entities} Across the phases, we discover individuals and categories~---~often with the help of \genai.  The common pattern is that we provide initial, `authoritative examples' (as in a seed set) request additional candidates from \genai{} (seeded entity discovery) and request classification (few-shot classification), followed by review and selection by us. This is often an iterative process. For example, rather than trying to enumerate all conceivable software engineering activities ourselves or mapping one specific resource to a formal taxonomy ourselves, we leverage \genai{} to collect types of activities at varying degrees of detail; see Sec.~\ref{S:phase8}. Review often relies on supporting Wikipedia pages for a given subject.
  \item \textbf{Properties} Properties were identified (`discovered') solely by the author, who took responsibility for shaping the emerging ontology in this respect. That is, the initial introduction of a property~---~formally by means of an OWL property declaration~---~is always a human-led initiative. In theory, we could ask \genai{} for additional proposals for properties; this may be a sensible thing to do, as we assume that the underlying model knows of ontologies and can compare the emerging \fsl{} ontology with existing ontologies and relate to best practices. In practice, so far, we find it demanding enough to work through our own expectations regarding properties. Naming for properties can be difficult, also sometimes in combination with deciding on the primary direction of an invertible property. For example, we were often struggling with names such as `has...', `uses', `serves', `facilitates'. We used \genai{} to discuss naming questions on the grounds of specifying the intended semantics of the relationship. We could have searched the web or looked through ontologies for inspiration instead. Using \genai{} simply increased productivity. As the ontology increased in size, it helped to use \genai{} for assertion discovery (few-shot completion) based on examples that we provided and subject to review of suggested assertions. For example, a number of programming languages (with which we are familiar) were associated with programming concepts by \genai; see Sec.~\ref{S:phase7}.
  \item \textbf{Axioms} \emph{domain/range axioms} are part of the conception of property declarations, which is an author-conducted responsibility. Authoring \emph{class axioms} (other than subclassing, see `classification'), i.e., `DisjointClasses', etc., is also an author-conducted responsibility, as such axioms require very specific domain knowledge. Some of the more obvious \emph{property axioms} (e.g., `inverseOf') could be suggested by \genai, when encountering common scenarios (e.g., `IrreflexiveProperty' in the context of mereology).
  \item \textbf{Transformations} Throughout the phases, the emerging ontology was rarely edited `manually' and, if so, only locally (such as for naming and comments). All other steps of introducing new entities, properties, assertions were verbally described and realized as OWL-level transformations executed by means of \genai. The ability to leverage \genai{} here was massively helpful in improving productivity. The initiative regarding these transformations remained completely with the author. Validation was straightforward~---~diff-based and subject to exploration in Prot\'eg\'e.
\end{enumerate}
\end{enumerate}





\section{Results}
\label{S:results}

We will present quantitative and qualitative results phase by
phase. The intermediate stages of \fsl{} can be observed in its GitHub
repo; there is a designated overview for the first release covered by
the present
paper.\footnote{\url{https://github.com/softlang/fsl/tree/main/misc/1st_release}}
Where \genai{} was used to operationalize steps of some
phases, transcripts of the conversations are available from the
overview.


\subsection{\phaseI}
\label{S:result1}

Based on our understanding of the notion of foundations of software
languages, we assume certain `subject areas of interest'; see the
goal \goalintro{Subject Area Agreement}{SAA} (Sec.~\ref{S:phase1}); we
assign priorities to them to rank their relevance, but also
to align with our expertise; see Table~\ref{T:subjareas} for the result.


\begin{table}[!htbp]
    \scriptsize
  \begin{tabular}{|l|l|}\hline
    \textbf{Subject area to be sampled} & \textbf{Priority}\\\hline\hline
    Software Language Engineering (SLE) & High\\\hline
    Compiler Construction (CC)  & High\\\hline
    Programming Language Theory (PLT)  & High\\\hline
    Theoretical Computer Science (TCS)  & Medium\\\hline
    Software Engineering (SE) & Low\\\hline
    Formal Methods (FM) & Low\\\hline
    Knowledge Representation \& Reasoning (KR)  & Low\\\hline
    Computational Linguistics (CL) & Low\\\hline
    Natural Language Processing (NLP) & Low\\\hline
  \end{tabular}
  \caption{Subject areas of interest}
  \label{T:subjareas}
\end{table}


Accordingly, we define a seed set with entities for \fsl;
see the goal \goalintro{Seed-Set Agreement}{SSA}
(Sec.~\ref{S:phase1}); see Table~\ref{T:seedset} for the result.


\begin{table}[!htbp]
  \scriptsize
  \begin{tabular}{|l|l|l|}\hline
    \textbf{Foundational entity} & \textbf{Reason for inclusion} & \textbf{Area(s)}\\\hline\hline
    Context-free grammar (CFG) & A fundamental formalism in parsing & CC, SLE\\\hline
    Parsing expression grammar (PEG) & A more modern approach to parsing & CC, SLE\\\hline
    Extended Backus-Naur Form (EBNF) & A metasyntax (i.e., a syntax for syntaxes) & CC, SLE\\\hline
    Regular grammar & A fundamental formalism in parsing, in fact, scanning & CC, SLE\\\hline
    Attribute grammar & An important language-implementation technique & CC, SLE\\\hline
    Term-rewriting system & A formal and rule-based transformation approach & CC, SLE\\\hline
    Lambda calculus & All lambda calculi used across PLT, TCS, etc. & PLT\\\hline
    Untyped lambda calculus & The basic, untyped lambda calculus & PLT \\\hline
    Simply typed lambda calculus & The simply typed lambda calculus & PLT \\\hline
    System F & A lambda calculus with polymorphic types & PLT \\\hline
    Lambda cube & All typed lambda calculi & PLT \\\hline
    Denotational semantics & The denotational style of semantics specification & PLT \\\hline
    Operational semantics & The operational style of semantics specification & PLT \\\hline
    Axiomatic semantics & The axiomatic style of semantics specification & PLT \\\hline
    Process calculus & All calculi for modelling concurrent systems & PLT, TCS\\\hline
    Communicating sequential processes (CSP) & A concrete calculus on concurrency & PLT, TCS\\\hline
    Calculus of communicating systems (CCS) & A concrete calculus on concurrency & PLT, TCS\\\hline
    UML state machine & A modeling language for finite automata & FM, SE\\\hline
    Hoare logic & A prominent approach for program verification & FM\\\hline
    Description logic & A prominent family of logics for ontologies & KR\\\hline
    Dependency grammar & Grammatical theories based on dependency & CL, NLP \\\hline
  \end{tabular}
  \caption{Seed set for FSL}
  \label{T:seedset}
\end{table}


The selection is biased by our background and expertise, but the
sampling is relatively broad because of the need to cover diverse
subject areas of interest; the higher the priority of a subject area,
the more corresponding entities there are in the seed set. The seed set will
effectively serve as a focus group for validation in all
subsequent phases.

All entities were linked to Wikipedia (according to goal \goaltla{KRL}
as of Sec.~\ref{S:phase1}), that is, we favor Wikipedia for systematic
linkage of external knowledge resources for now. A common discussion
topic in ontology engineering with regard to linking to Wikipedia,
Wikidata, or DBpedia is which of these sources to use and what type of
link to use (`sameAs' or `seeAlso' or `primaryTopic' or `page'). At
the risk of sounding unorthodox, we decided in favor of Wikipedia
because of how important human validation and therefore understanding
the narrative of Wikipedia pages is. Also, we were looking into the
common `sameAs' and `seeAlso' challenge and decided eventually to
prefer FOAF-style `isPrimaryTopicOf' and `page', which are also
prepared for the identity issues on Wikipedia. We also allow links
with anchors for subpages.

\begin{figure}[!htbp]
\small
  \begin{center}
    \includegraphics[width=0.9\textwidth]{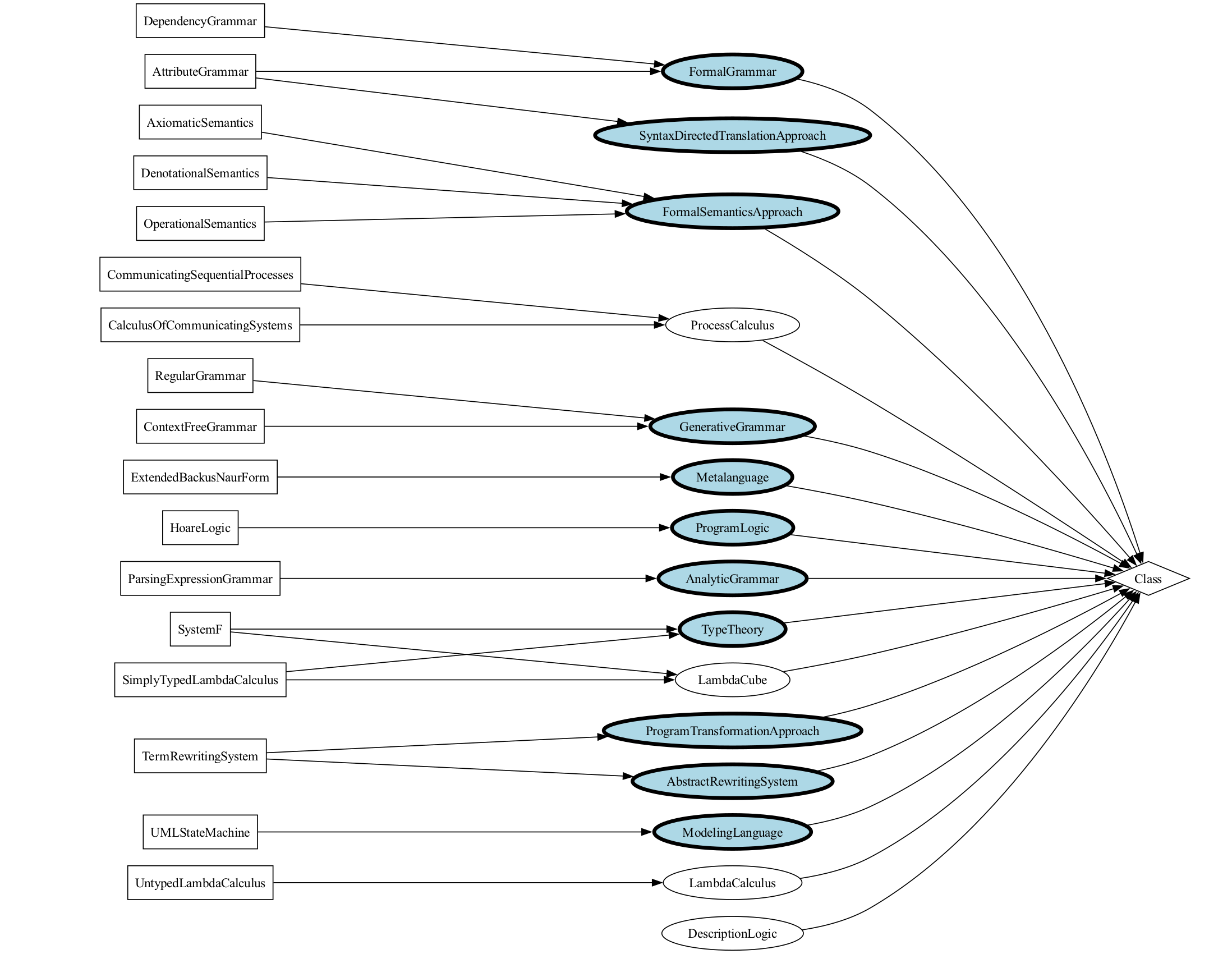}
  \end{center}

Legend: rectangles for individuals; ellipses for categories. Nodes
  added in this phase are highlighted to visualize the delta between
  this phase and the seed set.

\medskip
    \caption{Result of \phaseII}
    \label{F:phase2}
\end{figure}


\begin{figure}[!htbp]
\small
  \begin{center}
    \includegraphics[width=0.9\textwidth]{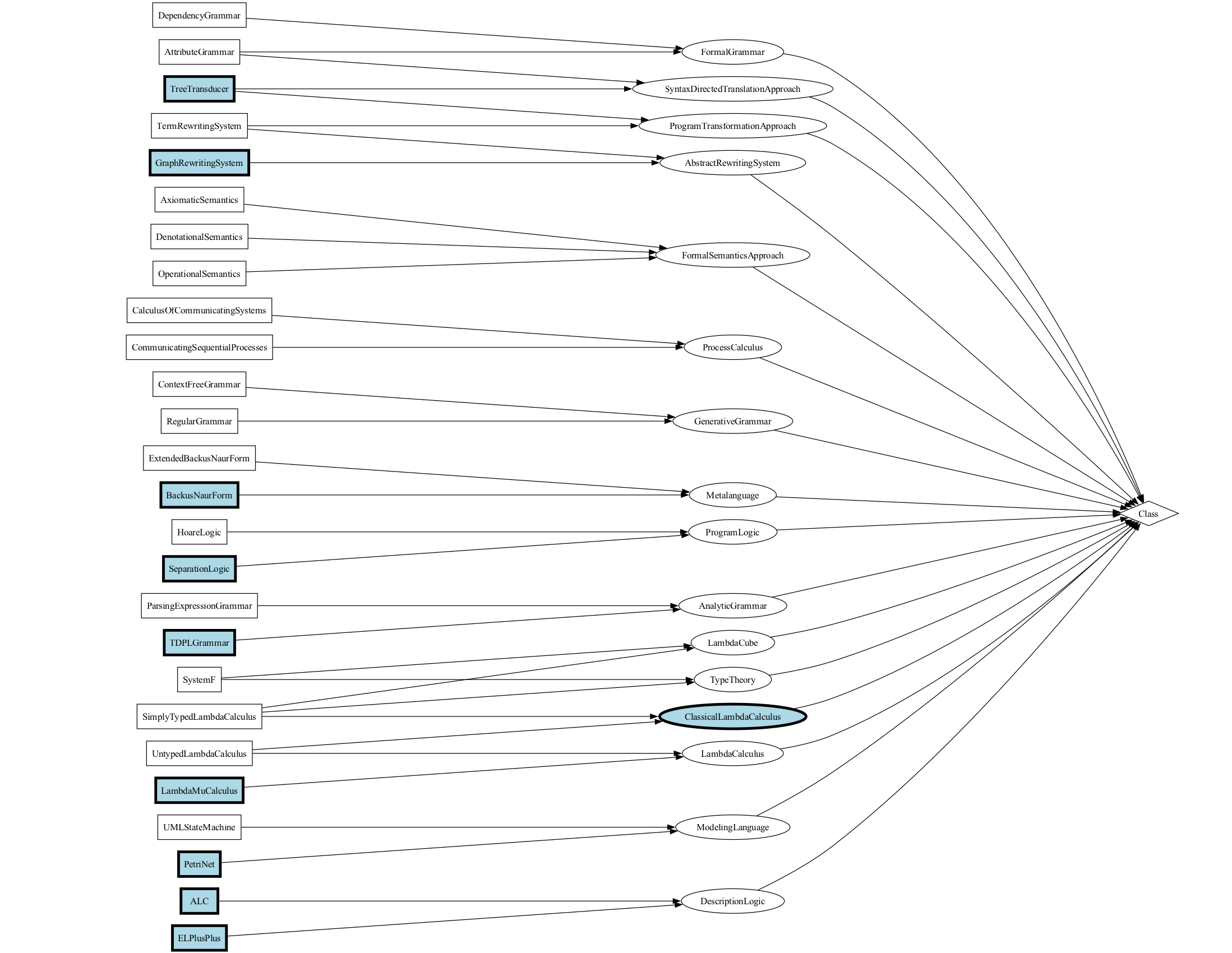}
  \end{center}

Legend: rectangles for individuals; ellipses for categories. Added
  categories and individuals are highlighted to visualize the delta
  between this phase and the previous phase.

\medskip
  \caption{Result of \phaseIII}
    \label{F:phase3}
\end{figure}

  
\subsection{\phaseII}
\label{S:result2}

The main goal of this phase was \goalintro{Individual/Category
  Distinction}{ICD} (Sec.~\ref{S:phase2}); see Fig.~\ref{F:phase2} for
the resulting hierarchy. As one can observe, some of the entities from
the seed set are set up to function as categories, whereas most
entities from the seed set are considered individuals. In order to
properly categorize all of the individuals, we had to include a number
of categories. In this paper, we did not (plan to) discover additional
individuals. It is also worth noting that we have a flat classification
hierarchy, that is, there is no subclassing yet between the categories.


\subsection{\phaseIII}
\label{S:result3}

The main goals of this phase were \goalintro{Double Instantiation
  Enforcement}{DIE} and \goalintro{Double Subclassing
  Enforcement}{DSE} (Sec.~\ref{S:phase3}); see Fig.~\ref{F:phase3} for
the resulting hierarchy. As one can observe, we mainly added instances
in response to \goaltla{DIE}.  As our classificatiobn hierarchy is
flat, there was nothing to be done with regard to
\goaltla{DSE}. Clearly, this property should be monitored throughout
subsequent expansion. However, we added a new category for `Classical
Lambda calculus' to better deal with the overloaded meaning of `Lambda
calculus'.


\begin{sidewaystable}[!htbp]
\small
\noindent
\begin{filecontents*}{snippets/properties.csv}
Property;Type;Domain;Range;Label;Comment;Inverse;Assertions
hasArea;O;Entity;SubjectArea;has area;Relates an entity to an important subject area of interest;isAreaOf;422
nativelySupports;O;LanguageEntity;LanguageConcept;Natively supports;Relates a language to a concept that it supports natively;;379
hasSpace;O;Entity;TechnologicalSpace;has space;Relates an entity to a technological space to which it characteristically belongs;isSpaceOf;186
supports;O;LanguageEntity;LanguageConcept;Supports;Relates a language to a concept that it supports;;126
targets;O;ToolEntity;LanguageEntity;targets;Relates a tool entity to a language entity that it targets;isTargetedBy;39
hasTool;O;TechnologicalSpace;ToolEntity;has tool;Relates a technological space to a kind of tool that it is concerned with;isToolOf;38
processes;O;ToolEntity;ArtifactEntity;processes;Relates a tool entity to an artifact entity that it characteristically processes;isProcessedBy;29
uses;O;Entity;Entity;uses;Relates an ontology entity to another ontology entity that it uses conceptually, methodologically, or technically;isUsedBy;27
elementOf;O;ArtifactEntity;LanguageEntity;element of;Relates an artifact to a language in the set-theoretic sense;hasElement;26
hasSupportingLanguage;O;EngineeringActivity;LanguageEntity;has supporting language;Relates a software engineering activity to a language kind that is characteristically used to perform or express the activity;;23
hasPrimaryComponent;O;CompositeConcept;LanguageConcept;Has primary component;Relates a composite language concept to a (primary) component;;23
hasSecondaryComponent;O;CompositeConcept;LanguageConcept;Has secondary component;Relates a composite language concept to a (secondary) component;;22
hasArtifact;O;TechnologicalSpace;ArtifactEntity;has artifact;Relates a technological space to a kind of artifact that it is concerned with;isArtifactOf;21
hasInputArtifact;O;;ArtifactEntity;has input artifact;Relates an activity or a tool to an artifact kind that it typically requires or consumes as input;;19
hasOutputArtifact;O;;ArtifactEntity;has output artifact;Relates an activity or a tool to an artifact kind that it typically produces as output;;16
hasSupportingTool;O;EngineeringActivity;ToolEntity;has supporting tool;Relates a software engineering activity to a tool kind that characteristically supports carrying out the activity;;15
metamodelingPolicy;A;;;Metamodeling policy;Documents ontology-wide expectations for explicit punning-based typing of classes and individuals;;14
assertedForConcept;O;LanguageSupportAssertion;LanguageConcept;Asserted for concept;Relates a reified assertion about language concepts with the concept;;13
assertedForLanguage;O;LanguageSupportAssertion;LanguageEntity;Asserted for language;Relates a reified assertion about language concepts with the language;;13
hasDominantForm;O;LanguageSupportAssertion;LanguageConcept;Has dominant form;Relates a language concept to a dominated form;;13
hasSupportKind;O;LanguageSupportAssertion;LanguageSupportKind;Has support kind;Relates a reified assertion about language concepts with the type of support;;13
serves;O;Entity;MethodologicalApproach;serves;Relates an entity to a methodological approach that it serves, realizes, supports, or instantiates in an important way;isServedBy;12
hasPrimaryMechanism;O;LanguageSupportAssertion;ImplementationMechanismConcept;Has primary mechanism;Relates a reified assertion about language concepts with the primary mechanism for realization;;12
hasBibTeX;D;Entity;string;has BibTeX;Relates an entity to the BibTeX key of a bibliographic entry that documents, discusses, or supports it;;11
commentingPolicy;A;;;Commenting policy;Documents the ontology-wide expectation regarding commenting;;9
formattingPolicy;A;;;Formatting policy;Documents the ontology-wide expectation regarding formatting;;9
linkingPolicy;A;;;Linking policy;Documents ontology-wide expectations for linking classes and individuals to descriptive web pages;;9
hasAffectedArtifact;O;;ArtifactEntity;has affected artifact;Relates an activity or a tool to an artifact kind that it typically affects (thereby complementing in- and output);;7
conformsTo;O;ArtifactEntity;ArtifactEntity;conforms to;Relates an artifact to another artifact in the generalized sense of models conforming to metamodels;classifies;6
classifies;O;ArtifactEntity;ArtifactEntity;classifies;Relates an artifact to another artifact in the generalized sense of metamodels classifying models;conformsTo;5
extends;O;SoftwareLanguage;SoftwareLanguage;extends;Relates a software language to another software language that it extends or generalizes;isExtendedBy;3
standardlySupports;O;LanguageEntity;LanguageConcept;Standardly supports;Relates a language to a concept that it supports standardly;;3
isSpecifiedBy;O;FormalEntity;MethodologicalApproach;is specified by;Relates an entity to a methodological approach used to formally specify or define its behavior or meaning;specifies;2
defines;O;ArtifactEntity;LanguageEntity;defines;Relates an artifact to a language in the sense of a definition (specification);isDefinedBy;1
\end{filecontents*}
\pgfplotstabletypeset[
    col sep=semicolon,
    string type,
    columns={Property,Comment},
    sort,
    sort key=Property,
    sort cmp={string <},
    columns/Property/.style={string type,column type=l},
    columns/Comment/.style={column type=l},
    every head row/.style={
        before row=\toprule,
        after row=\midrule
    },
    every last row/.style={
        after row=\bottomrule
    }
]{snippets/properties.csv}
\caption{\fsl{} properties~---~Overview I/II}
\label{T:properties1}
\end{sidewaystable}


\begin{sidewaystable}[!htbp]
\small
\noindent
\pgfplotstabletypeset[
    col sep=semicolon,
    string type,
    columns={Property,Assertions,Type,Domain,Range,Inverse},
    sort,
    sort key=Assertions,
    sort cmp={int >},
    columns/Property/.style={string type,column type=l},
    columns/Assertions/.style={fixed, precision=0,column type=l,column name=\#\,Assertions},
    columns/Type/.style={column type=l},
    columns/Domain/.style={column type=l},
    columns/Range/.style={column type=l},
    columns/Inverse/.style={column type=l},
    every head row/.style={
        before row=\toprule,
        after row=\midrule
    },
    every last row/.style={
        after row=\bottomrule
    }
]{snippets/properties.csv}
\begin{center}
Legend: Type O for object properties, A for annotation properties, D for datatype properties.
\end{center}
\caption{\fsl{} properties~---~Overview II/II}
\label{T:properties2}
\end{sidewaystable}


\subsection{\phaseIV}
\label{S:result4}

This phase aimed at the introduction of an initial set of
properties. Property discovery was continued in phases~5, 7, 8, as new
types of entities were included. In Table~\ref{T:properties1}, all
properties (not just those initially introduced in this phase) are
listed with an explanatory comment to get a first impression of the
kind of properties discovered. In Table~\ref{T:properties2},
additional metadata about the properties is provided. From the number
of assertions, we can infer the modest size of \fsl{} in its
first release, calling for future growth (`completion').


\begin{table}[!htbp]
\small
\noindent
\begin{center}
\begin{filecontents*}{snippets/spaces.csv}
Metric,Count
instances,13
subclasses,0
spaces_with_languages,13
languages_with_spaces,101
spaces_with_tools,13
tools_with_spaces,26
spaces_with_artifacts,13
artifacts_with_spaces,21
\end{filecontents*}
\pgfplotstabletypeset[
    col sep=comma,
    string type,
    columns/Metric/.style={verb string type, column type=l},
    columns/Count/.style={column type=l, column name=Count},
    every head row/.style={
        before row=\toprule,
        after row=\midrule
    },
    every last row/.style={
        after row=\bottomrule
    }
]{snippets/spaces.csv}
\end{center}

Legend: `instances' for the number
    of technological spaces; `subclasses` being 0 to mean that no
    subclassing on spaces is attempted; `spaces\_with\_$\ldots$' for the
    number of spaces with assertions to a given entity type; likewise
    for `$\ldots$\_with\_spaces'.

\medskip
    
  \caption{Quantities for \phaseV}
  \label{T:spaces}
\end{table}


\subsection{\phaseV}
\label{S:result5}

All goals of this phase (Sec.~\ref{S:phase5} are directly aimed at
the addition of the notion of technological spaces to
\fsl. Table~\ref{T:spaces} summarizes the result~---~essentially by
showing how (types of) languages, tools, and artifacts associate with
technological spaces. We refer back to Fig.~\ref{F:modelware} for
illustration where we see indeed all such different types of entities
which engage in `conformsTo', `processes', `serves' et al.\ relationships.

We also experimented with a variation on \goalintro{Knowledge Resource
  Linkage}{KRL} in this phase in so far that scholarly references had
to be provided for technological spaces and the related notion of
megamodeling as well as key concepts involved in megamodeling, notably
conformance and membership; see Table~\ref{T:hasBibTeX} for a first
attempt. We are not satisified with this status. In particular, there
is no place yet in \fsl, where the notion of megamodeling would be
directly reified; we use `MegamodelArtifact' as a proxy for now.


\begin{table}[!htbp]
\small
\noindent
\begin{center}
\begin{filecontents*}{snippets/references.csv}
Entity,Reference
MegamodelArtifact,FavreLaemmelVaranovich2012LinguisticArchitecture
MegamodelArtifact,FavreNguyen2004MegamodelEvolution
Grammarware,KlintLaemmelVerhoef2005Grammarware
Modelware,Bezivin2006MDETechnicalSpace
TechnologicalSpace,KurtevBezivinAksit2002TechnologicalSpaces
conformsTo,AtkinsonKuehne2003MetamodelingFoundation
conformsTo,AtkinsonGerbig2016OntologicalClassification
conformsTo,DegueuleEtAl2017SafeModelPolymorphism
conformsTo,RossiniDeLaraGuerraEtAl2014DeepMetamodelling
conformsTo,Favre2005MetaPyramids
elementOf,Favre2005MetaPyramids
\end{filecontents*}
\pgfplotstabletypeset[
    col sep=comma,
    string type,
    columns/Entity/.style={column type=l},
    columns/Reference/.style={
        column type=l,
        assign cell content/.code={
            \pgfkeyssetvalue{/pgfplots/table/@cell content}{\cite{##1}}%
        }
    },
    every head row/.style={
        before row=\toprule,
        after row=\midrule
    },
    every last row/.style={
        after row=\bottomrule
    }
]{snippets/references.csv}
\end{center}

\medskip
    
  \caption{hasBibTeX assertions for \phaseV}
  \label{T:hasBibTeX}
\end{table}


As an aside, the phase-by-phase discussion of results focuses on
the main goals per phase, while in reality extra modifications were
performed on the emerging ontology, as it could be observed from the
intermediate versions and, in many cases, also from the transcripts of
the LLM conversations. For example, in this phase, we actually i)
established the ultimate name of \fsl; ii) we introduced `extends' for
some prior uses of `hasPart'; iii) we performed another renaming on
what (later) would be renamed again to the final category `FormalEntity'.



\subsection{\phaseVI}
\label{S:result6}

The main goal of this phase was the modularization of \fsl. To this
end, we would first address the goal \goalintro{Top Level
  Refactoring}{TLR} (Sec.~\ref{S:phase6}) to review and improve the
top-level of the entity-type hierarchy. The top-level plus some
selected branches were previously shown in
Fig.~\ref{F:entity_types}. Quantities for the top-level entity types
are shown in Table~\ref{T:entity_types}~---~as of the end of
preparing the first release. Because of
punning/metamodeling~---~\goalintro{OWL 2 Punning Enforcement}{OPE}
(Sec.~\ref{S:phase3})~---~we count subclasses also as `entities'.


\begin{table}[!htbp]
\small
\noindent
\begin{center}
\begin{filecontents*}{snippets/entities.csv}
Entity type,Number of entities,Number of instances,Number of subclasses
Language entity,103,69,34
Formal entity,70,44,26
Conceptual entity,280,70,129
Tool entity,55,23,32
Artifact entity,28,0,28
Property entity,51,51,0
\end{filecontents*}
\pgfplotstabletypeset[
    col sep=comma,
    string type,
    columns/Entity type/.style={column type=l},
    columns/Number of entities/.style={column type=l,column name=\#\,Entities},
    columns/Number of instances/.style={column type=l,column name=\#\,Instances},
    columns/Number of subclasses/.style={column type=l,column name=\#\,Subclasses},
    every head row/.style={
        before row=\toprule,
        after row=\midrule
    },
    every last row/.style={
        after row=\bottomrule
    }
]{snippets/entities.csv}
\end{center}
\caption{Quantities for the top-level entity types of \fsl}
\label{T:entity_types}
\end{table}


\begin{table}[!htbp]
\small
\noindent
\begin{center}
\begin{filecontents*}{snippets/modules.csv}
Ontology,URI,Purpose
fsl,http://www.softlang.org/ontologies/fsl,FSL as the union of imports
tbox,http://www.softlang.org/ontologies/tbox,Shared TBox of FSL
fe,http://www.softlang.org/ontologies/fe,ABox for formal entities
pe,http://www.softlang.org/ontologies/pe,ABox for programming languages and their concepts
le,http://www.softlang.org/ontologies/le,ABox for (other) software languages and their concepts
te,http://www.softlang.org/ontologies/te,ABox for software tools
ae,http://www.softlang.org/ontologies/ae,ABox for software artifacts
ce,http://www.softlang.org/ontologies/ce,ABox for conceptual entities
ie,http://www.softlang.org/ontologies/ie,ABox for pending issues
\end{filecontents*}
\pgfplotstabletypeset[
    col sep=comma,
    string type,
    columns={Ontology,Purpose},
    columns/Ontology/.style={verb string type, column type=l},
    columns/Purpose/.style={verb string type, column type=l},
    every head row/.style={
        before row=\toprule,
        after row=\midrule
    },
    every last row/.style={
        after row=\bottomrule
    }
]{snippets/modules.csv}
\end{center}
\caption{\fsl's modules}
\label{T:modules}
\end{table}


\begin{figure}[!htbp]
  \centering
  \begin{tabular}{p{0.58\textwidth} p{0.36\textwidth}}
    \includegraphics[width=0.57\textwidth]{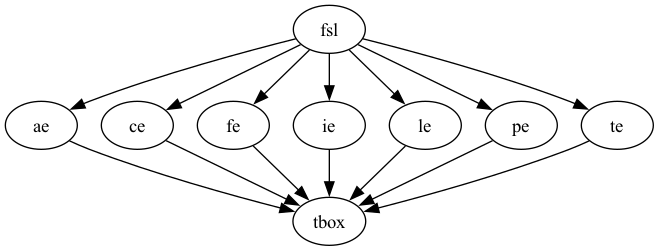} &
    \includegraphics[width=0.35\textwidth]{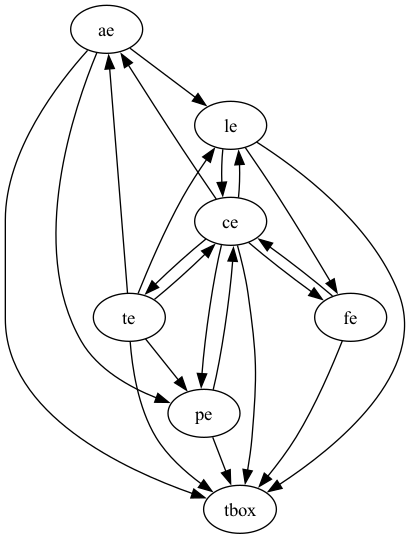} \\
    \centering Imports & \centering Namespaces \\
  \end{tabular}
  \caption{Dependencies for \fsl's modules}
  \label{F:modules}
\end{figure}


In addressing the goal \goalintro{Modules per Entity Types}{MET}, we
obtained the modules listed in Table~\ref{T:modules} with the
dependencies show in Fig.~\ref{F:modules}~---~as of the end of
preparing the first release; we should note that the module `pe' was
only introduced in \phaseVII{} (Sec.~\ref{S:phase8}). The import
dependencies are in line with the idea that the tbox is completely
centralized in one module; all other modules are aboxes~---~with
subclass- and possibly class-related constraints though. The namespace
dependencies are a consequence of how subjects of a given namespace
associate with objects of other namespaces.

As an aside, we realized only in this phase that we had forgotten
an important subject area of interest: MDE (model-driven
engineering). This turned out not to be a problem because relevant
entities were discovered during \phaseV{} (Sec.~\ref{S:phase5}), as
`modelware' is an important technological space which closely aligns
with MDE.


\subsection{\phaseVII}
\label{S:result7}

In accordance with the goal \goalintro{Programming Paradigm
  Coverage}{PPC} (Sec.~\ref{S:phase7}), there are corresponding
concepts included in \fsl; refer back to
Fig.~\ref{F:language_concepts} for an illustration.  Quantitive
results regarding the other goals (\goalintro{PL Concept
  Coverage}{PCC}, \goalintro{Programming Language Coverage}{PLC}, and
\goalintro{SL Concept Coverage}{SCC}) (Sec.~\ref{S:phase7}) are
summarized in Table~\ref{T:language_concepts}. For example, we see
that a significant number of concepts are `in use' and both
programming languages and software languages of other types are
annotated with concepts.


\begin{table}[!htbp]
\small
\begin{center}
\begin{filecontents*}{snippets/language_concepts.csv}
Metric,Count
concepts,156
instances,125
subclasses,31
used_instances,118
used_subclasses,31
properties,8
software_languages,47
programming_languages,27
\end{filecontents*}
\pgfplotstabletypeset[
    col sep=comma,
   string type,
    columns/Metric/.style={verb string type, column type=l},
    columns/Count/.style={column type=l, column name=Count},
    every head row/.style={
        before row=\toprule,
        after row=\midrule
    },
    every last row/.style={
        after row=\bottomrule
    }
]{snippets/language_concepts.csv}
\end{center}

Legend: `concepts' for the number of concepts with the breakdown
`instances' and `subclasses' (all being individuals due to punning);
`used\_instances' and `used\_subclasses' for clarifying what concepts
are actually exercised in ontological assertions (thereby also
providing evidence of punning); `properties' for the number of
ontological properties exercised in assertions with concept
involvement; `software\_languages' for the number of actual software
languages with concept-based assertions with `programming\_languages'
for the fraction thereof being specifically programming languages.

\medskip

\caption{Quantities for \phaseVII}
\label{T:language_concepts}
\end{table}


\begin{table}[!htbp]
\small
\noindent
\begin{center}
\begin{filecontents*}{snippets/se_activities.csv}
Metric,Count
instances,0
immediate_subclasses,10
nonimmediate_subclasses,80
used_nonimmediate_subclasses,17
properties,6
language_instances,0
language_subclasses,10
tool_instances,0
tool_subclasses,9
artifacts,6
\end{filecontents*}
\pgfplotstabletypeset[
    col sep=comma,
    string type,
    columns/Metric/.style={verb string type, column type=l},
    columns/Count/.style={column type=l, column name=Count},
    every head row/.style={
        before row=\toprule,
        after row=\midrule
    },
    every last row/.style={
        after row=\bottomrule
    }
]{snippets/se_activities.csv}
\end{center}

Legend: `instances' (being 0) as a testament to the fact that \fsl{}
does not capture actual activities in actual projects (but
applications of \fsl{} very well may do so); `immediate\_subclasses'
(being 10) illustrates the coverage of major activities in the
software lifecycle with `nonimmediate\_subclasses' being much larger
serving as an indication of the nuanced types of SE activities
covered of which (see `used\_nonimmediate\_subclasses') however only
relatively few are already exercised; `properties' for the number of
ontological properties exercised in assertions with SE activity
involvement; `language\_$\ldots$' and `tools\_$ldots$' for the numbers
of languages and tools exercised in assertions with SE activity
involvement where the zeroes at the `$\ldots$\_instances' level
indicate that punning is at play such that SE activity types associate
with types of languages and tools~---~rather than specific languages
and tools; `artifacts' for the number of artifact types because \fsl{}
does not capture actual artifacts in actual projects or repositories
and general activity types are also just concerned with types of artifacts.

\medskip

\caption{Quantities for \phaseVIII}
\label{T:se_activities}
\end{table}


\subsection{\phaseVIII}
\label{S:result8}

In Table~\ref{T:se_activities}, quantitative results regarding the
goals of this phase~---~\goalintro{Software Lifecycle Coverage}{SLC}
and \goalintro{Technological Space Coverage}{TSC}
(Sec.~\ref{S:phase8})~---~are summarized. The top-level of the
resulting hierarchy of SE activities exactly corresponds to the
software lifecycle. All subclasses are just more nuanced. The activity
types are not yet exercised much. Technological space coverage is best
claimed by identifying types of languages, tools, and artifacts being
associated with SE activity types; coverage is slim here at this
stage.


\subsection{\phaseIX}
\label{S:result9}

Reviewing was obviously a permanent task during the development
process for the first release of \fsl, but we concentrated a more
systematic effort in the last phase. We will walk through the
corresponding goals as of Sec.~\ref{S:phase9} now.


\begin{table}[!htbp]
\small
\noindent
\begin{center}
\begin{filecontents*}{snippets/constraints.csv}
Property,Uses
disjointWith,19
inverseOf,28
minQualifiedCardinality,4
onClass,4
onProperty,6
someValuesFrom,2
unionOf,19
\end{filecontents*}
\pgfplotstabletypeset[
    col sep=comma,
    string type,
    columns/Property/.style={verb string type, column type=l},
    columns/Uses/.style={column type=l, column name=\#\,Uses},
    every head row/.style={
        before row=\toprule,
        after row=\midrule
    },
    every last row/.style={
        after row=\bottomrule
    }
]{snippets/constraints.csv}
\end{center}
\caption{OWL constraints (\phaseIX)}
\label{T:constraints}
\end{table}


The goal \goalintro{Reasoning-Based Consistency}{RBC}
(Sec.~\ref{S:phase9}) does not require any documentation other than
that the HermiT Reasoner completes fine on the first release of
\fsl. However, let us also summarize the constraints for the first
release of \fsl; see Table~\ref{T:constraints}. One may draw the
conclusion that \fsl{} (as of the first release) is relatively
underspecified; see the discussion of future work (Sec.~\ref{S:evolve}).


\begin{table}[!htbp]
\scriptsize
\fbox{%
\begin{minipage}{0.96\linewidth}
\begin{description}
\item[ClassDeclarationsMustHaveCommentShape] Class-declarations must
  be `documented' by means of a comment.
\item[PropertyDeclarationsMustHaveCommentShape] Ditto for property declarations.
\item[ClassDeclarationsMustHaveFoafLinkShape] Class-declarations must
  be linked with the `real world' by a FOAF link.
\item[PropertyDeclarationsMustHaveFoafLinkShape] Ditto for property declarations.
\item[MetamodelingShapes] The different types of entities (especially
  in terms of their subclasses) must conform to a schema of punning/metamodeling.
\item[TechnologySpacesMustHaveConformsToShape] Technological spaces
  must engage in at least one conformance relationship involving two
  types of space-related artifacts which in turn relate to each other
  in the sense of conformance.
\item[TechnologySpacesMustBeSpecifiedShape] Technological spaces must
  be `specified' sufficiently in terms of related languages, tools,
  and artifacts.
\item[EngineeringActivitiesMustBeSpecifiedShape] Software engineering
  activities must be `specified' in terms of artifacts for input,
  output, or otherwise consumed or affected, also in terms of
  supporting languages and tools.
\item[MethodologicalApproachesMustBeSpecifiedShape] Methodological
  approaches must be `specified' in the sense that they are connected
  to other types of entities, most notably in the sense that some
  formal entity or language serves an approach.
\end{description}
\end{minipage}%
}
\caption{SHACL shapes (\phaseIX)}
\label{T:shacl}
\end{table}


The goal \goalintro{Shape-Based Validation}{SBV} (Sec.~\ref{S:phase9})
is addressed in Table~\ref{T:shacl} where we summarize the shapes that
readily constrain the first release of \fsl; see GitHub for
details. There are definitely some areas for improvement. For example,
\fsl{} uses punning in a manner that would better be documented and
enforced more strongly.


\begin{table}[!htbp]
\small
\begin{filecontents*}{snippets/queries.csv}
Query theme, Queries, Purpose
phase2, 1, Initial FSL taxonomy tree of phase 2
phase3, 1, FSL taxonomy tree for delta between phases 2 and 3
entities, 1, FSL entity type-usage metrics
properties, 1, FSL property-usage metrics
spaces, 8, Specification aspects for technological spaces in FSL
language_concepts, 8, Specification aspects for language concepts in FSL
se_activities, 10, Specification aspects for SE activities in FSL
modules, 2, Graphs for import and namespace dependencies of FSL
constraints, 1, Constraint usage metrics for FSL
references, 1, Use of scholarly references in FSL
\end{filecontents*}
\pgfplotstabletypeset[
    col sep=comma,
    string type,
    columns/Query theme/.style={verb string type, column type=l},
    columns/Queries/.style={column type=l, column name=\#\,Queries},
    columns/Purpose/.style={verb string type, column type=l},
    every head row/.style={
        before row=\toprule,
        after row=\midrule
    },
    every last row/.style={
        after row=\bottomrule
    }
]{snippets/queries.csv}
\caption{SPARQL queries (\phaseIX)}
\label{T:sparql}
\end{table}


The goal \goalintro{Query-Based Reporting}{QBR} (Sec.~\ref{S:phase9})
has been essentially demonstrated throughout Sec.~\ref{S:results}
because all tables and figures are based on systematically authored
queries; see Table~\ref{T:sparql} for a summary.


The goal \goalintro{Text-Based Transformation}{TBT}
(Sec.~\ref{S:phase9}) is more like a guideline for all phases and
certainly for the final review phase so that transformations should be
described and recorded, thereby resulting in better
transparency/reproducibility. In this manner, we are committing to a
basic principle that is established all across software engineering
and computer science; in the more schema-based context at hand, we
also have contributed to such a transformation-based development of
specifications~---~especially in the context of
grammars~\cite{LaemmelV01-SPE,
  LaemmelV01-IEEE,Laemmel01-FME,KlintLV05} (even though typically
based on specifications in an appropriate transformation language
rather than text-based descriptions, but the latter is now practical
with \genai{} support).


\begin{figure}[!htbp]
\footnotesize
\fbox{%
\begin{minipage}{0.96\linewidth}
\begin{verbatim}
@prefix ... -- omitted

<http://www.softlang.org/ontologies/ie> a owl:Ontology ;
    ..,
    rdfs:comment "Sub-ontology / ABox for issue entities. ..."@en ;
    owl:imports <http://www.softlang.org/ontologies/tbox> .

:SupportsNameClashIssue a tbox:IssueEntity ;
    tbox:target tbox:supports ;
    tbox:target ce:supports ;
    tbox:critique "There are two different types of 'supports' in use.
      tbox:supports relates tools to languages, whereas ce:supports
      relates languages to concepts. They cannot be usefully combined
      under a shared abstraction, unless we are willing to have longer
      property names that, perhaps, include the range into the property
      name for disambiguation."@en ;
    tbox:suggestion "Rename tbox:supports to tbox:targets,
        since a tool may be said to target a language."@en .

:LanguageConceptsNamespaceIssue a tbox:IssueEntity ;
    tbox:target ce:LanguageConcept ;
    tbox:target ce:CompositeConcept ;
    tbox:target ce:LanguageSupportKind ;
    tbox:target ce:LanguageSupportAssertion ;
    tbox:target ce:LanguageConceptApplicability ;
    tbox:target ce:ImplementationMechanismConcept ;
    tbox:target ce:supports ;
    tbox:target ce:nativelySupports ;
    tbox:target ce:standardlySupports ;
    tbox:target ce:assertedForConcept ;
    tbox:target ce:assertedForLanguage ;
    tbox:target ce:hasDominantForm ;
    tbox:target ce:hasPrimaryMechanism ;
    tbox:target ce:hasApplicability ;
    tbox:target ce:hasComponent ;
    tbox:target ce:hasPrimaryComponent ;
    tbox:target ce:hasSecondaryComponent ;
    tbox:target ce:hasSupportKind ;
    tbox:critique "Several schema-level language-concept classes and properties are
      currently declared in the ce module, although they belong to the shared tbox
      vocabulary. This was previously considered acceptable, but the module boundary
      should now be cleaned up."@en ;
    tbox:suggestion "Move all targeted class and property declarations from ce
      to tbox, change their namespace from ce to tbox, and update all references
      accordingly."@en ;
    tbox:resolveAfter :SupportsNameClashIssue .

:ModuleImportIssue a tbox:IssueEntity ;
    tbox:target <http://www.softlang.org/ontologies/ae> ;
    tbox:target <http://www.softlang.org/ontologies/le> ;
    tbox:target <http://www.softlang.org/ontologies/pe> ;
    tbox:target <http://www.softlang.org/ontologies/te> ;
    tbox:critique "Ideally, all modules other than tbox and fsl should only import tbox.
      This is violated for no good reason."@en ;
    tbox:suggestion "Remove the violating imports from the targeted modules."@en ;
    tbox:resolveAfter :LanguageConceptsNamespaceIssue .
\end{verbatim}
\end{minipage}%
}
\caption{Illustrative sets of issues fixed in \phaseIX}
\label{F:ie}
\end{figure}


Throughout the development of the first release of \fsl, we described
several transformations in the LLM-based conversations, as evident
from the corresponding transcripts, but in the review phase, we
deliberately switched to explicit capture of the transformation in an
ontological format; there were three batches of issues; see
Fig.~\ref{F:ie} for the last one which happens to be relatively small,
thus fit for presentation. The idea is that `target' refers to the
primary resource that requires transformation, `critique' formulates
the problem with the present ontology; `suggestion' describes the
actual transformation; `resolveAfter' helps with ordering of issue
resolution. The set of issues at hand is obviously concerned with
cleaning up some related property naming and entity placement (in
namespaces) and module-import dependencies.


\begin{table}[!htbp]
\small
\fbox{%
\begin{minipage}{0.96\linewidth}
\begin{multicols}{2}
\begin{Verbatim}
conformsToIssue
describesIssue
hasArtifactIssue
hasReferenceGeneralizationIssue
hasSpaceGeneralizationIssue
hasToolIssue
isAreaOfNormalizationIssue
isProcessedByToolIssue
servesDomainIssue
supportsNormalizationIssue
tboxCommentingIssue
tboxFormattingIssue
ArtifactsDisjointnessIssue
ConceptsDisjointnessIssue
ConformanceLiteratureIssue
FormalEntitiesDisjointnessIssue
FormalSystemsDisjointnessIssue
FoundationalToFormalIssue
LanguagesDisjointnessIssue
LinkingPolicyIssue
MakePropertiesEntitiesIssue
MigrateToBibTeXIssue
MissingPoliciesIssue
PropertyGraphIssue1
PropertyGraphIssue2
ToolsDisjointnessIssue
tboxDependsOnCEIssue
tboxDependsOnFEIssue
LanguageConceptsNamespaceIssue
ModuleImportIssue
SupportsNameClashIssue
\end{Verbatim}
\end{multicols}
\end{minipage}%
}
\caption{Issues covered along the first release of \fsl}
\label{T:issues}
\end{table}


In Table~\ref{T:issues}, all issues~---~ontologically coded as
illustrated above~---~are listed; see GitHub for details. The issues
from Fig.~\ref{F:ie} are also included in the list. See also the
discussion of future work for some outstanding
issues (Sec.~\ref{S:evolve}) and the need to use CI/CD for a more
systematic process of issue tracking and resolution (Sec.~\ref{S:cicd}). 

\section{Discussion}
\label{S:discuss}

We will now summarize our findings and suggest directions for future research.


\subsection{Summary of findings}

We draw the following conclusions from the conducted research:

\begin{description}

\item[An ontological definition of the software languages field] Up to
  now, software languages and their engineering have been
  conceptualized only informally and pragmatically. \fsl{} provides a
  formalized conceptualization in terms of taxonomic and ontological
  building blocks. \fsl{} classifies software languages and connects
  them with language tools, language concepts, the underlying
  formalisms, software engineering activities and technological spaces
  (as to how they involve languages).

\item[A software language-focused instantiation of ontology
  engineering] Every domain requires specific efforts for
  ontology engineering. We identified the specific aspects required
  to cover software languages in the \fsl{} context. That is, we
  included phases of covering language concepts, the underlying
  formalisms, software engineering activities and technological
  spaces. All these types of entities are now integrated in \fsl{}
  and released for use, review, and enhancement on GitHub. 

\item[Proven usefulness of \genai{} in the software languages field]
  We received significant help from \genai{} regarding
  entity discovery and classification, ontology completion, and
  linkage of knowledge resources. \genai{} was also specifically
  useful for ontology manipulation based on textual specifications of
  intended transformations that were found to be reliably executable
  via \genai.

\end{description}


\subsection{Future research directions}
\label{S:future}

There are three major directions: i) evolution of \fsl{} in a
relatively standard manner; ii) leverage of CI/CD for the continuous
improvement of \fsl; iii) pairing of the \fsl{} ontology with suitable
chrestomathic efforts. We will discuss these three directions in turn.


\subsubsection{Evolution of \fsl}
\label{S:evolve}

There are many relatively obvious issues regarding the necessary and
unsurprising evolution of FSL. We have filed corresponding issues on
GitHub.\footnote{\url{https://github.com/softlang/fsl/issues?q=is\%3Aissue\%20label\%3A\%22past\%20V1\%22}}
We group these issues here as follows:

\begin{description}
\item[Validation] Some issues are about SHACL-based hardening of validation.
\item[Completion] Other issues are about further ontology completion, as some
  incompleteness scenarios are readily known, for example, not all
  technological spaces are sufficiently specified.
\item[Exception elimination] In fact, several issues deal with
  validation and completion at once~---~in the sense that the existing
  SHACL shapes make exceptions due to observed incompleteness so that
  validation passes for now.
\item[Consistency] Other issues are concerned with ontology
  consistency regarding structure, constraints, and documentation. For
  example, the systematic and relatively exhaustive use of
  disjointness constraints and investigation of the use of
  definitional axioms (`EquivalentClass') are part of this
  theme. Also, the class hierarchy needs some work. For example,
  `tbox:ConceptualEntity` combines very different types of entities,
  which creates some confusion for domain/range properties. Also, the
  use of policies (for documentation, metamodeling and linking) is
  somewhat ad hoc at this point. Also, we use domain/range+cardinality
  constraints in some areas, where these are rather validation-related
  expectations, as the OWA does not serve our expectations here; we
  were using these constraints to pass intentions to the LLM.
\item[Bibliography] We also aim at systematic bibliographic grounding.
\end{description}


\subsubsection{CI/CD for \fsl}
\label{S:cicd}

CI/CD (Continuous Integration and Continuous Delivery/Deployment) for
ontologies would certainly include aspects such as issue templates,
SHACL and SPARQL tests, competency-question regression tests, reasoner
checks, documentation generation, and defined and checked release
criteria. CI/CD is a developing field in ontology engineering and
knowledge management. Before we explain how we specifically want to
apply CI/CD to \fsl{} and how we expect to make a contribution,
we summarize the state of the art.

ROBOT~\cite{Jackson2019ROBOT} is a tool for automating ontology
workflows; while it is not framed explicitly in the CI/CD context,
the approach covers some of the aspects listed above. ODK (Ontology
Development Kit)~\cite{Matentzoglu2022ODK} is a toolkit for building,
maintaining, and standardizing biomedical ontologies; it standardizes
automatically executable workflows for quality control, dependency
management, and releases. (Software-like deployment~---~the D in
CI/CD~---~is often seen as release engineering in the ontology
context.) OntoFlow~\cite{Dziwis2022OntoFlow} provides a general
workflow-based view on ontology
engineering. OnToology~\cite{Alobaid2019OnToology} automates
documentation, evaluation, releasing, and versioning for Git-based
ontology development; this is a good example of relatively early
CI-style support automation, before the CI/CD label was used more
explicitly. By contrast, Ontolo-CI~\cite{Publio2022OntoloCI} is an
explicitly CI-labeled approach which leverages GitHub Actions for RDF
and ontology-adjacent validation using
ShEx. ACIMOV~\cite{Hannou2023ACIMOV} is a methodology which thus goes
beyond tool support and proposes a development method based on
modularity, Git workflows, automated syntactic/semantic checks,
documentation generation, and
publication. SAREF~\cite{Lefrancois2023SAREFPipeline} is a strong
example of current CI/CD state of the art for ontology engineering; it
covers compliance checks, dependency handling, SHACL-based
verification, and automated portal generation/publication.


Even with knowledge resources that are not strict ontologies,
DevOps-/CI/CD-like methods have been explored. For instance, the newer
DBpedia Release Cycle~\cite{Hofer2020DBpediaRelease} concerns a major
knowledge graph rather than ontology engineering per se, but the
approach is strongly CI/CD-shaped: automated release workflow, testing
methodology, regular releases, productivity/agility, and
maintainability. The idea of using DevOps principles for semantic data
quality dates back much longer~\cite{Meissner2016DevOpsRDF}. In our
own work on online software chrestomathies with semantic
labeling~\cite{FavreLSV12,FavreLLSV12,LaemmelSV13,LaemmelLSV14,SchaussEtAl17},
we also explored simple forms of quality control and dependency
management regarding, for example, consistency of Wiki content with underlying
source code as well as correct use of a RDFS-like vocabulary applied
for feature modeling or architectural modeling. 

Much of this CI/CD work for ontologies should be relatively
straightforward to apply to \fsl, for example, GitHub Actions,
workflows for SHACL and reasoning, template-based issue generation,
competency-question regression tests, release management,
documentation generation, etc.~---~we are currently working on such
aspects. Given the early state of \fsl's development and the
particularities of \fsl's domain, a few particular concerns are worth
mentioning, which in some cases indeed require research: a) \fsl{} and
Wikipedia are linked in a challenging manner, which requires constant
monitoring regarding correctness, drift, and discovery in combination
with semi-automatic workflows for corrective measures; b) \fsl{} is
highly incomplete in terms of available candidate ABox knowledge such
that prioritized discovery would be needed to arrive at a manageable
process for improvement with the human in the loop; c) \fsl{} is
inspired by much pre-ontological knowledge, as we have reported,
but the alignment with those resources should be controlled more
formally, subject to appropriate semi-formalization of the underlying
resources and semi-automated integration thereof.


\subsubsection{Pairing of \fsl{} with chrestomathies}
\label{S:chresto}

As discussed in Sec.~\ref{S:slepreonto}, we consider software
chrestomathies~\cite{Laemmel15} and megamodels in the software (language)
engineering context~\cite{Laemmel17} important pre-ontological
knowledge resources informing ontology engineering for
\fsl. Ultimately, \fsl{} as an ontology should converge with a
suitable collection of megamodels and a suitable chrestomathy. Let us
sketch such a future pairing:

\begin{description}
\item[\yas] \yas\footnote{\url{https://github.com/softlang/yas/}} (`Yet
  another SLR (Software Language Repository)') is a rich
  collection (chrestomathy) of language-related artifacts serving the
  demonstration of various language-related aspects (such as
  specification, implementation, analysis, and transformation). \yas{}
  is the repository providing the foundation of the author's software
  languages book~\cite{SoftLangBook18}. Interestingly, \yas{}
  internally uses two types of megamodeling languages for multi-level
  relationship maintenance~\cite{Laemmel17}. It would be relatively
  obvious to incorporate a referencing layer into \fsl{} to associate
  with \yas{} for demonstrative purposes. We actually plan such a
  layer in the ongoing effort of preparing the next edition of the
  book, as the book is also meant to benefit from \fsl{} and to cover
  ontology engineering as a subject (in the sense of ontologies being
  considered software languages, too).
\item[101companies] The 101companies chrestomathy on software languages,
  technologies, and
  concepts~\cite{FavreLSV12,FavreLLSV12,LaemmelLSV14} and a related
  metaprogramming-focused spinoff~\cite{LaemmelLSV14,SchaussEtAl17}
  suggest another referencing layer for \fsl{} in principle, but
  101companies is no longer actively maintained and does not cover
  current software (language) engineering practice. However, the
  underlying idea of implementing a small software system
  idiomatically and in alignment with an underlying feature model
  across many languages and technologies remains a viable path for a
  knowledge resource. Coverage of software engineering activities was
  limited for 101companies; broader coverage would be natural for
  \fsl.
\item[Formal entities] \fsl{} suggests another chrestomathic
  direction: formal entities in the context of software
  languages. Such a chrestomathy would cover formal systems and other
  formal entities relevant in the context of software languages. In
  the most direct manner, all formalisms covered in \fsl{} should be
  properly demonstrated. In a less obvious manner, the existing
  category of programming (or software) language concepts could be
  illustrated by appropriately labeled language definitions and uses
  of formalisms as well as sample programs.
\end{description}

\section{Conclusions}
\label{S:concl}

We have presented the initial development of \fsl~---~an emerging
ontology for the foundations of software languages. \fsl{} covers
software languages, their categories, tools related to them,
concepts underlying those languages, software engineering activities
involving languages, and technological spaces
contextualizing language usage.

\fsl{} is already useful for teaching and research in the author's
context in several ways. First, \fsl{} now provides a knowledge
resource to inject structure into courses on software language
engineering, programming language theory, and ontology
engineering. Second, \fsl{} will be leveraged to cover the
software language category of ontologies in the next edition of the
Software Languages Book (i.e., the 2nd edition following
\cite{SoftLangBook18}). Some of these efforts will advance \fsl;
community contributions will be very helpful.



%


\vspace{6pt}

\reftitle{References}


\bibliography{bibs/oe.bib,bibs/misc.bib,bibs/softlangbooke1.bib,bibs/cicd.bib,bibs/fsl.bib}

%


\PublishersNote{}
\end{document}